\documentclass[traditabstract]{aa}

\usepackage{graphicx}
\pdfoutput=1
\usepackage{amsmath}
\usepackage{txfonts}
\usepackage{natbib}
\usepackage[usenames,dvipsnames]{color}
\usepackage{enumerate}
\newcommand{\binc}{\texttt{binary\_c}}

\newcommand{\Mprim}{M_{1,\mathrm{i}}}

\newcommand{\Msec}{M_{2,\mathrm{i}}}
\newcommand{\Mpmz}{M_{\mathrm{PMZ}}}
\newcommand{\sepi}{a_{\mathrm{i}}}
\newcommand{\logg}{\log_{10} g}
\newcommand{\loggunits}{\log_{10}(g/\mathrm{cm}\,\mathrm{s}^{-2})}
\newcommand{\Teff}{T_{\mathrm{eff}}}
\newcommand{\Pin}{P_{\mathrm{i}}}
\newcommand{\Pf}{P_{\mathrm{f}}}
\newcommand{\Porb}{P_{\mathrm{orb}}}

\newcommand{\Vmin}{V_{\mathrm{min}}}
\newcommand{\Vmax}{V_{\mathrm{max}}}

\newcommand{\cm}{\mathrm{cm}}
\newcommand{\Msun}{M_{\odot}}
\newcommand{\Rsun}{R_{\odot}}
\newcommand{\Lsun}{L_{\odot}}
\newcommand{\Hy}{\mathrm{H}}
\newcommand{\C}{\mathrm{C}}
\newcommand{\Cth}{^{13}\mathrm{C}}

\newcommand{\Fe}{\mathrm{Fe}}
\newcommand{\Ge}{\mathrm{Ge}}

\newcommand{\Ba}{\mathrm{Ba}}

\newcommand{\Pb}{\mathrm{Pb}}
\newcommand{\Eu}{\mathrm{Eu}}
\newcommand{\X}{\mathrm{X}}
\newcommand{\Y}{\mathrm{Y}}

\newcommand{\fC}{F_{\mathrm{C}}}
\newcommand{\fCdwa}{F_{\mathrm{C,dwa}}}
\newcommand{\fCTO}{F_{\mathrm{C,TO}}}
\newcommand{\fCgia}{F_{\mathrm{C,gia}}}
\newcommand{\fdwaC}{F_{\mathrm{dwa, CEMP}}}
\newcommand{\fTOC}{F_{\mathrm{TO, CEMP}}}
\newcommand{\fgiaC}{F_{\mathrm{gia, CEMP}}}

\newcommand{\IL}{\mathrm{I_{L05}}}
\newcommand{\IK}{\mathrm{I_{K01}}}
\newcommand{\SDM}{\mathrm{S_{DM}}}

\newcommand{\Cq}{\mathrm{C_q}}

\begin{document}


   \title{Modelling the observed properties of carbon-enhanced \\metal-poor stars using binary population synthesis}
   \titlerunning{Modelling the observed properties of CEMP stars using binary population synthesis}

   \author{C. Abate
          \inst{1,2}
          \and
          O. R. Pols \inst{2} 
          \and
          R.J. Stancliffe\inst{1}
          \and
          R. G. Izzard \inst{1,3}
		  \and
		  A. I. Karakas \inst{4}
		  \and
		  T. C. Beers \inst{5}
		  \and
		  Y. S. Lee \inst{6}
          }

   \institute{Argelander Institut f\"ur Astronomie, Auf dem H\"ugel 71, D-53121 Bonn, Germany\\\email{cabate@uni-bonn.de}
         \and
        Department of Astrophysics/IMAPP, Radboud University Nijmegen, P.O. Box 9010, 6500 GL Nijmegen, The Netherlands
         \and
        Institute of Astronomy, Madingley Road, Cambridge CB3 0HA, United Kingdom
         \and
        Research School of Astronomy \& Astrophysics, Mount Stromlo Observatory, Weston Creek ACT 2611, Australia
         \and
		Department of Physics and JINA-CEE, 225 Nieuwland Science Hall, Notre Dame, IN 46656, USA
         \and
        Deptartment of Astronomy \& Space Science, Chungnam National University, Daejeon 305-764, Republic of Korea
             }

   \date{Received ...; accepted ...}
 

  \abstract
   {
   {The stellar population in the Galactic halo is characterised by a
   large fraction of carbon-enhanced metal-poor (CEMP) stars. Most CEMP
   stars have enhanced abundances of $s$-process elements (CEMP-$s$
   stars), and some of these are also enriched in $r$-process elements
   (CEMP-$s/r$ stars). One formation scenario proposed for CEMP stars
   invokes wind mass transfer in the past from a thermally-pulsing
   asymptotic giant branch (AGB) primary star to a less massive
   companion star which is presently observed.
}
   {
   In this work we generate synthetic populations of binary stars at
   metallicity $Z=0.0001$ ($[\Fe/\Hy]\approx-2.3$), with the aim of
   reproducing the observed fraction of CEMP stars in the halo. In addition, we aim
   to constrain our model of the wind mass-transfer process, in
   particular the wind-accretion efficiency and angular-momentum loss,
   and investigate under which conditions our model populations reproduce
   observed distributions of element abundances.
   }
   {
   We compare the CEMP fractions determined from our synthetic
   populations and the abundance distributions of many elements with
   observations. Several physical parameters of the binary stellar
   population of the halo are uncertain, in particular the initial mass
   function, the mass-ratio distribution, the orbital-period
   distribution,
   and the binary fraction. We vary the assumptions in our model about
   these parameters, as well as the wind mass-transfer process, and study the
   consequent variations of our synthetic CEMP population.
   }
   {The CEMP fractions calculated in our synthetic populations vary
   between $7\%$ and $17\%$, a range consistent with the CEMP fractions
   among very metal-poor stars recently derived from the SDSS/SEGUE data
   sample. %
   The resulting fractions are more than a factor of three higher
   than those determined with default assumptions in previous
   population-synthesis studies, which typically underestimated the
   observed CEMP fraction.
   We find that most CEMP stars in our simulations are
   formed in binary systems with periods longer than $10,\!000$ days.
   Few CEMP stars have measured orbital periods, but all that do have
   periods up to a few thousand days. Our results are consistent only if
   this small subpopulation represents the short-period tail of the
   underlying period distribution. The results of our comparison between
   the modelled and observed abundance distributions are significantly
   different for CEMP-$s/r$ stars and for CEMP-$s$ stars that are not
   strongly enriched in $r$-process elements. For these stars, our
   simulations qualitatively reproduce the observed distributions of
   carbon, sodium, and heavy elements such as strontium, barium,
   europium, and lead. Contrarily, for CEMP-$s/r$ stars our model
   cannot reproduce the large abundances of neutron-rich elements such
   as barium, europium, and lead. This result is consistent with previous
   studies, and suggests that CEMP-$s/r$ stars experienced a
   different nucleosynthesis history to CEMP-$s$~stars.
   }
   }

\nopagebreak   
\maketitle


\section{Introduction}
\label{intro}

The population of old metal-poor stars observed in the Galactic halo
carries information about the conditions under which the Milky Way was
formed. In the past three decades, the HK survey \cite[][]{Beers1985,
BeersAJ1992}, the Hamburg/ESO survey \cite[HES;][]{Christlieb2001,
Christlieb2008}, and the SDSS/SEGUE survey \cite[][]{York2000, Gunn2006,
Yanny2009} have collected spectra of a large sample of very metal-poor
stars (VMP, that is, stars with abundances of iron\footnote{The
relative abundance of two elements X and Y is $[\X/\Y] =
\log_{10}(N_{\X}/N_{\Y}) - \log_{10}(N_{\X}/N_{\Y})_{\odot}$, where
$N_{\X}$ and $N_{\Y}$ refer to the number density of atoms of
X and Y, respectively, and $\odot$ indicates the solar values.}
$[\Fe/\Hy]\lesssim-2$ ), and have made it possible to investigate their
dynamical and chemical properties. These studies reveal a large
proportion of carbon-enhanced metal-poor (CEMP) stars. 
According to the original nomenclature proposed by 
\cite{BeersChristlieb2005}, CEMP stars have $[\C/\Fe]>+1$. 
More recently a number of authors have adopted the criterion
$[\C/\Fe]>+0.7$ to define CEMP stars \cite[e.g.][]{Aoki2007, Lee2013,
Yong2013III}. The fraction of CEMP stars among metal-poor stars is
observed to rise with increasing distance from the Galactic plane
\cite[][]{Frebel2006, Carollo2012}. A strong increase in the cumulative
fraction of CEMP stars with decreasing metallicity has
been confirmed by many studies, from about $3\%$ at $[\Fe/\Hy]<-1$
to around $75\%$ at $[\Fe/\Hy]<-4$
\cite[e.g.][]{Cohen2005, Marsteller2005, Frebel2006, Lucatello2006, Carollo2012,
Aoki2013a, Lee2013, Yong2013III, Placco2014}.

CEMP stars are usually classified into four categories according to the
observed abundances of barium and europium, two elements associated with
the {\it slow} ($s$-) and {\it rapid} ($r$-) neutron-capture process,
respectively.

\begin{itemize}
\item[$\bullet$] {\it CEMP-$s$ stars} exhibit 
enhanced abundances of $s$-elements, and are defined by $[\Ba/\Fe]>+0.5$
and $[\Ba/\Eu]>0$. CEMP-$s$ stars account for at least $80\%$ of all
observed CEMP stars \cite[][]{Aoki2007}. A number of authors adopt
slightly different criteria, e.g. $[\Ba/\Fe] > +1$ and/or $[\Ba/\Eu]>+0.5$
\cite[][]{BeersChristlieb2005, Jonsell2006, Masseron2010}.
\item[$\bullet$] {\it CEMP-$s/r$ stars}%
\footnote{The original nomenclature for these stars is ``CEMP-$r/s$''; 
we employ CEMP-$s/r$ for consistency with Papers~I and II.} %
are CEMP-$s$ stars that also enriched in europium and $r$-elements, $[\Eu/\Fe]>+1$.
\item[$\bullet$] {\it CEMP-$r$ stars} are a rare CEMP 
subclass enriched in elements produced purely by the $r$-process, and are
defined by the criteria $[\Eu/\Fe]>+1$ and $[\Ba/\Eu]<0$.
\item[$\bullet$] {\it CEMP-no stars} do not exhibit peculiar
enhancements in the elements heavier than iron \cite[][]{Aoki2002}.
\end{itemize}

The formation scenario of CEMP stars is still uncertain, and different
categories of CEMP stars are likely to have different formation channels.
Several mechanisms have been proposed to explain the large carbon abundances, for example:
($a$) the gas cloud in which these stars were formed was already
enriched in carbon produced by zero-metallicity and/or rapidly rotating
stars \cite[][]{Mackey2003, Meynet2006, Meynet2010, Maeder2014} or
expelled by the faint supernovae associated with the first generation of
stars \cite[][]{Umeda2003, Umeda2005, Nomoto2013}; and ($b$) the
carbon-enhanced material was accreted from the envelope of a
thermally-pulsing asymptotic giant branch (AGB) primary star in a binary
system onto the presently observed low-mass companion. The mass-transfer
scenario provides a natural explanation for the chemical abundances of
CEMP-$s$ stars because carbon and $s$-elements are produced during AGB
nucleosynthesis, and the observed fraction of CEMP-$s$ stars with a
binary companion is consistent with the hypothesis of all CEMP-$s$ stars being in
binary systems \cite[][]{Lucatello2006, Starkenburg2014}. By contrast,
many CEMP-no stars show no indication of a binary companion
(\citealp{Norris2013-2}; Hansen et al., {\it in prep.}), and they exhibit
low abundances of $s$-elements, whereas large enhancements of
$s$-elements are normally produced in AGB stellar models, except in some
models of mass $M_*>3\,\Msun$. The origin of CEMP-$s/r$ stars is also an
open issue. Current stellar models indicate that in AGB stars the
density of free neutrons is not sufficiently large for the $r$-process
to take place, hence the enhancement of $r$-elements in CEMP-$s/r$ stars
remains unexplained. 

A number of authors have simulated populations of very metal-poor stars
with the aim of reproducing the observed fraction of CEMP stars. The
largest uncertainties in these models include the amount of carbon and
$s$-elements produced by AGB nucleosynthesis as a function of the
stellar mass and metallicity, the shape of the initial mass function
(IMF) of the early Galaxy, the binary fraction, the efficiency of the
mass-transfer process, and the range of separations in which it occurs. 
\cite{Lucatello2005a} and \cite{Komiya2007} argue that the IMF in the early
Galaxy was weighted towards intermediate-mass stars to account for the
large fraction of CEMP stars at low metallicity. \cite{Suda2013} and
\cite{Lee2014} suggest that the transition between the early (top-heavy)
and the present-day IMF occurred over the metallicity range between
$[\Fe/\Hy]=-2.5$ and $[\Fe/\Hy]=-1.5$. However, \cite{Pols2012} show
that, with an IMF biased towards intermediate-mass stars, many
nitrogen-enhanced metal-poor stars are produced, whereas very few are
observed.

\cite{Izzard2009} and \cite{Abate2013} model 
the CEMP population at metallicity $Z=10^{-4}$ (that is
$[\Fe/\Hy]\approx-2.3$) with their population-synthesis models, adopting
the solar-neighbourhood IMF proposed by \cite{Kroupa1993}, and varying a
set of uncertain physical parameters related to the mechanism of mass
transfer, nucleosynthesis, and mixing processes. They typically find CEMP
fractions between $2\%$ and $4\%$, while the observed values found by
different authors at $[\Fe/\Hy]\le-2$ vary between $9\%$ and $25\%$
\cite[][]{Marsteller2005, Frebel2006, Lucatello2006, Lee2013}. However,
their AGB evolution models are based on detailed models that do not
allow for third dredge-up in AGB stars of mass below approximately
$1.2\,\Msun$. To increase the CEMP/VMP fraction, \cite{Izzard2009} and
\cite{Abate2013} artificially allow for third dredge-up in AGB stars
down to mass $0.8\,\Msun$ in their models (as proposed by \citealp{Izzard2004}
using independent arguments), and show that in this way it is
possible to raise the modelled CEMP/VMP ratio to $\gtrsim10\%$.

The most recent detailed models of AGB evolution and nucleosynthesis of
\cite{Stancliffe2008}, \cite{Karakas2010} and \cite{Lugaro2012} find
third dredge-up in stars of mass down to $0.9\,\Msun$ at metallicity
$Z=10^{-4}$. In the models of \cite{Karakas2010} and \cite{Lugaro2012},
the nucleosynthesis products of stars are computed on a grid of $16$
initial masses in the range $[0.9,\,6]\,\Msun$, taking into account
$320$ isotopes from $^1\Hy$ up to $^{210}\mathrm{Po}$. In our recent
work \cite[][Paper I and II hereafter]{Abate2015-1,Abate2015-2} we
include the results of these detailed calculations in our model of
binary evolution and nucleosynthesis, and study in detail the chemical
compositions for a sample of $67$ CEMP-$s$ stars. We compare the surface
abundances of each CEMP-$s$ star with a grid of model binary stars, and
determine the best fit to the observed abundances. From this comparison
we conclude that our models reproduce reasonably well the chemical
properties of CEMP-$s$ stars with $[\Eu/\Fe]<+1$ (for brevity,
``$r$-normal CEMP-$s$'' stars hereafter), but not CEMP-$s/r$ stars,
possibly indicating different nucleosynthesis histories for the two
categories. Also, our models typically predict orbital periods longer
than those observed, suggesting that wind mass transfer should be more
efficient at close separations.

In this paper we extend our recent work on CEMP stars, and analyse the
properties of the population of CEMP stars at $[\Fe/\Hy]\approx-2.3$ as
a whole, through a comparison with synthetic populations. Our purpose is
to answer the following questions:

\begin{enumerate}
\item Is it possible to reproduce the observed CEMP/VMP 
ratio with our model of binary population synthesis that includes the
latest detailed AGB models of \cite{Karakas2010} and \cite{Lugaro2012}?
\item Can we constrain our model of the wind mass-transfer process, 
and in particular the wind-accretion efficiency and angular-momentum
loss?
\item Under which conditions does our model reproduce 
the observed abundance distributions of elements associated with AGB
nucleosynthesis?
\end{enumerate}

The paper is organised as follows. In Sect. \ref{models} we describe the
most important characteristics of our model and the selection criteria
that we adopt to compare the results of our synthetic population with
the observations. In Sect. \ref{results} we present the results of our
models, which are further discussed in Sect. \ref{discussion}.
Our conclusions are presented in Sect.~\ref{concl}.


\section{Models}
\label{models}

In this work we use the binary population synthesis code
\texttt{binary\_c/nucsyn} described by \cite{Izzard2004, Izzard2006,
Izzard2009} to simulate populations of stars at low metallicity, and
compare them with a sample of observed stars. In this section we
summarise the physical parameters of our model (Sect. \ref{physics}),
the assumptions made in our grid of binary models (Sect. \ref{grid}),
and describe the method used to select the simulated stars according to
their luminosity (Sect. \ref{method}).


\subsection{Input physics}
\label{physics}

In our default model set A we adopt the same input physics as in model
set A of Papers I and II. We briefly list our most important assumptions
below. Table \ref{tab:models} provides an overview of our alternative
model sets, in which we varied some of these assumptions.

\begin{itemize}
\item[$\bullet$] The \citet{Reimers75} equation multiplied by a factor
of $\eta=0.5$ is used to compute the wind mass-loss rate up to the AGB.
The prescription of \cite{VW93} describes mass loss during the AGB
phase, with minimum and maximum wind velocities
$v_{\mathrm{w}} = 5 \mathrm{~km\,s^{-1}}$ and $v_{\mathrm{w}} = 15
\mathrm{~km\,s^{-1}}$, respectively. 
\item[$\bullet$] We adopt the approximation of a spherically-symmetric 
wind (\citealt{Abate2013}, Eq. 4) to compute the angular momentum
carried away by the ejected material, and calculate the wind-accretion
rate according to a wind Roche-lobe overflow (WRLOF) model described by
Eq. (9) of \cite{Abate2013}. Our model sets B and C are as in
Paper II. In both model sets the expelled material carries away a
multiple $\gamma=2$ of the average specific orbital angular momentum of
the binary system. In model set B we adopt an enhanced
Bondi-Hoyle-Lyttleton (BHL) model of wind accretion computed with Eq.
(6) of \cite{BoffinJorissen1988} and $\alpha_{\mathrm{BHL}}=10$, whereas
in model set C our default WRLOF prescription is adopted (Paper~II).
\item[$\bullet$] The algorithms that we use to compute the 
nucleosynthesis of the star through the first and second dredge-ups are
based on the results of \cite{Karakas2002} and \cite{Karakas2007}; we
refer the interested reader to \cite{Izzard2004,Izzard2006,Izzard2009} for a
thorough description.
\item[$\bullet$] The amount of material dredged up during the AGB phase
is computed with algorithms tuned to reproduce the detailed models of
\cite{Karakas2010} at metallicity $Z=10^{-4}$, as described in Paper~I.
The chemical composition of the intershell region of the AGB star is
tabulated as a function of the mass of the star at the beginning of the
AGB phase, the evolution along the AGB, and the mass of the partial
mixing zone, $\Mpmz$, a free parameter of our model. In the detailed
AGB models of \cite{Karakas2010}, the partial mixing zone~(PMZ) is the
site in which protons from the envelope are partially mixed with
material of the intershell region. The protons react with the $^{12}\C$
nuclei and form a $\Cth$ pocket, in which free neutrons are produced
and become available for the $s$-process. We refer the reader to
\cite{Karakas2010} and \cite{Lugaro2012} for a detailed explanation of
the numerical treatment of the PMZ, and to Paper I for a description of
our implementation in our population-synthesis code.
\item[$\bullet$] We assume efficient thermohaline mixing, that is, 
the accreted material is instantaneously mixed with the accreting star.
The calculations of \cite{Stancliffe2007} suggest this approximation is
reasonable in many cases, even though \cite{Stancliffe2008} and
\cite{Stancliffe2009} show that other processes, such as gravitational
settling, may in some circumstances reduce the effect of thermohaline
mixing. To account for the possibility of inefficient thermohaline
mixing, in model set D the accreted material remains on the stellar
surface until mixed in by convection.
\item[$\bullet$] In our models we assume the metallicity of the detailed
 nucleosynthesis models by \cite{Lugaro2012}, i.e. $Z=10^{-4}$. We
adopt the same initial composition as Papers I and II based on the
chemical evolution models of \cite{Kobayashi2011} for the isotopes up to
$^{76}\Ge$. For heavier isotopes we assume the solar abundance
distribution of \cite{Asplund2009} scaled down to $Z=10^{-4}$. 
\end{itemize}

\begin{table}
\caption{Physical parameters adopted in our binary population models.
}
\label{tab:models}
\centering
\begin{tabular}{ c  c }
\hline
\hline
	model set &	Physical parameters	\\
			&	(differences from model set A)\\
\hline
A				&	Default \\
A1				&	$[\C/\Fe]>1.0$	\\
A4				&	As A1, $\loggunits<4.0$, no luminosity selection	\\[1ex]

B				&	Angular-momentum loss: $\Delta J/J=2\,(\Delta M/M)$ 	\\
				&	Wind-accretion efficiency: BHL, $\alpha_{\mathrm{BHL}}=10$\\
				&	(as model B of Paper I and II) 	\\[1ex]
C				&	Angular-momentum loss: $\Delta J/J=2\,(\Delta M/M)$ 	\\
				&	Wind-accretion efficiency: WRLOF	\\
				&	(as model C of Paper II)	\\[1ex]

D				&	No thermohaline mixing				\\

$\IK$			&	Multiple-part power-law IMF \cite[][]{Kroupa2001},			\\
				&	for $M>1\Msun\,$, IMF slope is less steep than default.			\\
$\IL$			&	Log-normal IMF \cite[][]{Lucatello2005a}, \\
				&	$\mu_{\log M}=0.79, \,\sigma_{\log M}=1.18$. \\

Q1				&	$\phi(q)\propto q^{~~~~}$				\\	
Qp4				&	$\phi(q)\propto q^{-0.4}$		\\

$\SDM$			&	Log-normal initial-period distribution		\\
				&	\cite[][]{DuquennoyMayor1991}, $\mu_{\log P}=4.8, \,\sigma_{\log P}=2.3$.		\\
S3				&	As $\SDM$ with $\mu_{\log P}=3.0$		\\

T8				&	$t_{\mathrm{min}}=8$ Gyr				\\
T12				&	$t_{\mathrm{min}}=12$ Gyr			\\
\hline
\end{tabular}
\end{table}
	

\subsection{Population synthesis}
\label{grid}

Each of our simulated populations consists of $N$ binary stars uniformly
distributed in the $\log_{10} M_1-M_2-\log_{10} a$ parameter space,
where $M_{1,2}$ are the masses of the primary and the secondary star,
respectively, and $a$ is the orbital separation. Seven different values
of the PMZ mass are taken into account. The adopted grid resolution is
$N=N_{\mathrm{M}1} \times N_{\mathrm{M}2}\times N_{a}\times
N_{\mathrm{PMZ}}$, where $N_{\mathrm{M}1}=100$, $N_{\mathrm{M}2}=32$,
$N_{a}=80$ and $N_{\mathrm{PMZ}}=7$. A finer grid with higher resolution
does not change our results.

The initial parameters vary in the following ranges:

\begin{itemize}
\item[$\bullet$] $\Mprim$ is in the interval $[0.5,\,8.0]\,\Msun$. 
Stars of mass below $0.5\,\Msun$ are not expected to be visible with the
current magnitude limitation of the observational surveys, as explained
in Sect. \ref{method}. Massive stars that end their evolution as
supernovae are not considered.
\item[$\bullet$] $\Msec$ is uniformly spaced in $[0.1,\,0.9]\,\Msun$. 
More massive stars have already evolved to a white dwarf after $10$ Gyr,
thus they are excluded because they do not contribute to the fraction of
CEMP stars. By definition, initially $\Msec\le\Mprim$, therefore the
initial mass ratio $q_{\mathrm{i}}=\Msec/\Mprim$ is always
$0<q_{\mathrm{i}}\le1$.
\item[$\bullet$] We consider circular orbits with $\sepi$ between $50$ and
$5\times10^6\Rsun$. Binary stars outside the considered range do not
become CEMP stars in our models, either because they are too close and they merge
when the primary star becomes a red giant, or because they are too wide to
interact. However, binary stars at closer separations are observed. For
example, \cite{Kouwenhoven2007} analyse the binary population of the
stellar association Scorpius OB2 in the range $[5,\,5\times10^6]\,
\Rsun$. We assume that all stars are formed in binary systems with
separations in this range. When calculating the CEMP fraction, we take
into account that our grid investigates a fraction of
the total population of binary stars. For example, with a flat
distribution in $\log_{10}\sepi$ our grid covers $83\%$ of the observed
range, thus the model CEMP fraction is multiplied by this factor. For
different $\sepi$ distributions we compute the corresponding
corrections. Binary systems with wider separations than $5\times10^6\,
\Rsun$ and single stars could in principle be taken into account by
decreasing the binary fraction.
\item[$\bullet$] When the stellar mass is $M_*<3\,\Msun$ we adopt
the following values of $\Mpmz$: $0$, $2\times10^{-4}$,
$5\times10^{-4}$, $6.6\times10^{-4}$, $1\times10^{-3}$,
$2\times10^{-3}$, $4\times10^{-3}\,\Msun$. $\Mpmz$ is always zero for
$M_*\ge3\,\Msun$.
\end{itemize}

Following \cite{Izzard2009} and \cite{Abate2013}, stars are counted according to the sum

\begin{eqnarray}
n_{\mathrm{type}} = S \sum_{M_{1,\mathrm{min}}}^{M_{1,\mathrm{max}}}\sum_{M_{2,\mathrm{min}}}^{M_{2,\mathrm{max}}}\sum_{a_{\mathrm{min}}}^{a_{\mathrm{max}}}\sum_{t_{\mathrm{min}}}^{t_{\mathrm{max}}} \delta_{\mathrm{type}}~ \Psi_{M_1,M_2,a}~ \delta M_1~\delta M_2~ \delta a~ \delta t~, 
\label{eq:nstars}
\end{eqnarray}
where
\begin{itemize}
\item $S$ is the star-formation rate, assumed to be constant for ages 
between $t_{\mathrm{min}}$ and $t_{\mathrm{max}}$.
\item $t_{\mathrm{min}}=10$ Gyr corresponds to the approximate minimum age of the Galactic halo,
while $t_{\mathrm{max}}=13.7$ Gyr is the age of the Universe; $\delta t$
is the timestep.
\item The volume unit in the parameter space is $\delta M_1 \cdot \delta M_2 \cdot \delta a$.
\item $\delta_{\mathrm{type}}=1$ when the star belongs to a specific
type, and zero otherwise. Model stars are classified as follows: VMP stars are
all stars older than $t_{\mathrm{min}}$ that have not become white
dwarfs; CEMP stars are VMP stars with $[\C/\Fe]> +0.7\,$; CEMP-$s$ stars
are CEMP stars with $[\Ba/\Fe]> +0.5$.
\item $\Psi_{M_1,M_2,a}$ is the inital distribution of $M_1$, $M_2$, and $a$.
We assume $\Psi$ to be separable,

\begin{equation}
\Psi = \Psi(M_1, M_2, a) = \psi(M_1)~\phi(M_2)~\chi(a)~.
\end{equation}

In our default model set A the primary mass distribution $\psi(\Mprim)$
is the IMF proposed by \cite{Kroupa1993}, the secondary mass
distribution $\phi(\Msec)$ is flat in $q_{\mathrm{i}}$, and the separation
distribution $ \chi(a)$ is flat in $\log a$.

\end{itemize}

Some of the assumed parameters in our model are uncertain and not
well-constrained by observational data, in particular: the IMF of the
primary star, the binary fraction, the distribution of the mass ratios
and separations in binary systems, and the age of the halo. To determine
how each of these uncertainties affects our results we modify our
default model set A by varying one parameter at a time. Table
\ref{tab:models} lists some of the model sets that we tested, and their
differences with respect to our default model set~A.


\subsection{Selection criteria}
\label{method}

In the papers by \cite{Izzard2009} and \cite{Abate2013}, the synthetic
stars are selected from the simulations when their surface gravities are
below the threshold $\loggunits=4.0$. This criterion essentially
restricts the analysis to giant stars, and is based on the implicit
assumption that all giant stars of the halo are visible. Here we replace
this criterion with a more realistic selection, based on the apparent
magnitudes of the stars, including in the analysis the main-sequence
stars that are sufficiently luminous to be visible.

To calculate the number of stars that are visible as a function of their
luminosity, $N(L)$, we follow the method adopted by \cite{Pim2014}. The
main steps are summarised as follows.

\begin{enumerate}
\item Our model computes the luminosity, $L$, the effective temperature, $\Teff$, 
and surface gravity, $\logg$, of each simulated star. The absolute
visual magnitude, $M_V$, is calculated from the luminosity as described
by \citet[][Eq. 10]{Torres2010}. The bolometric correction is computed
as a function of $\Teff$ and $\logg$ of the star, adopting the values
published by \citet[][Table 1]{Bessell1998}.
\item The apparent $V$ magnitude of a star is computed as a 
function of the distance from the Sun, $d$, and the Galactic latitude, $b$.
Similarly to \cite{Pim2014}, the extinction of stellar light is taken
into account according to the Galactic-dust distribution proposed by
\cite{Toonen2013} and the prescription of \cite{Sandage1972}. For
$b\neq0$, one has

\begin{equation}
V = M_V + 5\,(\log_{10}(d) - 1) + A_V(\infty)\tanh \left(\frac{d\sin b}{z_h}\right)~~,	\label{eq:V}
\end{equation}
where $z_h=120$ pc is the scale height of Galactic dust, and $A_V(\infty)$ is:
\begin{displaymath}
A_V(\infty) = \left\{	\begin{array}{lr}
0.165(\tan 50^{\circ} - \tan b) \csc b &	\hspace{2mm}\mathrm{if}~|b|<50^{\circ}~	\\
0	&	\hspace{2mm}\mathrm{if}~|b|\ge50^{\circ}.
\end{array}\right.
\end{displaymath}

\item For each simulated star of luminosity $L$ and corresponding absolute
magnitude $M_V$, we compute the volume in which the star has apparent
magnitude $\Vmin<V<\Vmax$. In our simulations, $\Vmin$ and $\Vmax$ are
set consistently with the detection limits of the observational data. We
compare our synthetic populations with two different data sets, which
have different magnitude ranges: 

\begin{itemize}
\item[$i$.] The observations of the SDSS/SEGUE survey (for
brevity, SEGUE), which include the effective temperatures, surface
gravities, and carbon abundances for about $13,\!000$ very metal-poor
stars with $-2.5\le[\Fe/\Hy]<-2.0$ observed at medium resolution
\cite[][]{Lee2013}. The formal $g$-magnitude range of the SEGUE survey
is $14<g<20.3$; however, \cite{Yanny2009} state that reliable
atmospheric parameter measurements are possible only within the range
$14<g<19$. We convert the limits in the $g$-band to $13.5<V<18.5$ with
the empirical photometric relation \cite[][]{Windhorst1991},

\begin{equation}
V = g - 0.03 - 0.42\,(g-r),
\end{equation}

\noindent where we adopt $(g-r)_{\mathrm{max}}=1.2$, as derived from Fig. 7 of \cite{Lee2013}.

\item[$ii$.] A sample of $378$ very metal-poor stars with $-2.8\le[\Fe/\Hy]\le-1.8$, 
both carbon-normal and carbon-enhanced, mostly based on the SAGA
database \cite[][last updates in January, 2015]{Suda2008, Suda2011}.
This is the sample adopted in Paper II, and includes high-resolution
measurements of the surface abundances of many elements. The apparent
$V$ magnitudes of the stars in this sample vary between $\Vmin=6$ and
$\Vmax=16.5$, hence we adopt this range in our simulations.

\end{itemize}
%

\item To determine $N(L, \Teff, \logg)$, that is, the number of visible stars with luminosity $L$, 
temperature $\Teff$, and surface gravity $\logg$, we integrate the
stellar density distribution of the Galactic halo over the volume
determined in the previous step. We assume a maximum distance of
$d=10^5$ pc. The stellar density distribution in the halo is
parameterised with the equation of an oblate spheroid in the reference
frame of the Galactic centre as prescribed by \cite{Helmi2008},

\begin{equation}
\rho(x,y,z)\propto r_{\odot}^{-n}\left(x^2+y^2+(\alpha z)^2\right)^{\frac{n}{2}},
\label{eq:rho}
\end{equation}
where $r_{\odot}=8$ kpc is the distance of the Sun from the Galactic
centre \cite[e.g.][]{Moni2012}, $\alpha$ is the minor-to-major axis
ratio, and $n$ is the exponent of the density profile. For $\alpha$ and $n$
we adopt the best-fit values determined by \cite{Juric2008} in their
three-dimensional density map of the Galactic halo: $\alpha=0.64$ and
$n=-2.8$.
\item To speed up the calculations we initially calculate $N(L, \Teff, \logg)$ 
for a grid of luminosities, temperatures, and gravities. The intervals of
these parameters in the grid are chosen to reproduce the ranges of
variation of these parameters in the binary stars of our simulations
(namely: $[-3,\,5]$, $[3000,\,8000]$ and $[-1,\,6]$ for
$\log_{10}(L/\Lsun)$, $\Teff/$~K and $\loggunits$, respectively). The
values of $N(L, \Teff, \logg)$ are stored in two tables, one for each
magnitude range considered.
\item Each star in our simulation is counted by multiplying the value
$n_{\mathrm{type}}$ computed with Eq. (\ref{eq:nstars}) by $N(L, \Teff,
\logg)$. The value of $N(L, \Teff, \logg)$ is determined by
interpolating $L, \Teff,$ and $\logg$ within our tables. This step is
repeated for all the synthetic populations generated with different
model sets.
\end{enumerate}

Finally, we note that in our models most horizontal-branch stars have
large effective temperatures, up to $10,\!000$ K. In the range
$-2.8\leq[\Fe/\Hy]\leq-1.8$, the SEGUE and SAGA databases contain only a
few horizontal-branch stars hotter than $6,\!000$ K, possibly because
hotter stars are selected against. In our simulations these stars are
typically not very numerous because the time spent on the horizontal
branch is short compared to the total stellar lifetime. However,
including them in our analysis modifies the model distribution of
surface gravities between approximately $2.5$ dex and $3.5$ dex, as
noted also by \cite{Izzard2009}. For this reason we exclude from our
simulations all stars with $\Teff>6,\!000$ K that have evolved further
than the turnoff. This selection has a negligible effect on the computed
fraction of CEMP stars.


\section{Results}
\label{results}
%


\subsection{Comparison with the results of \cite{Abate2013}.}
\label{compA13}

We now compare the CEMP-to-VMP fraction, $\fC$, computed with our default
model set A and model set $\mathrm{C_q}$ of \cite{Abate2013}. These two
model sets adopt the same assumptions about the distributions of initial
parameters, the wind-accretion efficiency, and the mechanism of
angular-momentum loss. However, the selection criteria used in these
model sets are different (Sect. \ref{method}). In model set A4 the same
definitions are used as in model set $\mathrm{C_q}$ to select VMP stars,
namely $\loggunits<4.0$ and $t_*\ge10$ Gyr, and the criterion
$[\C/\Fe]>+1$ is used to select CEMP stars. Because model set
$\mathrm{C_q}$ was computed assuming a range of initial separations
different from the present work, namely between $3\,\Rsun$ and $10^5\,
\Rsun$, the CEMP/VMP ratio of model $\mathrm{C_q}$ has been recalculated
to take into account this difference. 

\begin{table}[!t]
\caption{CEMP/VMP ratio, $\fC$, calculated with model sets A1 and A4 and with set $\Cq$ of \cite{Abate2013}. 
The errors convey only Poisson statistics. The SEGUE value of $\fC$ and the uncertainty are taken from \citet[Tab. 4]{Lee2013} for stars in the range $-2.5\le[\Fe/\Hy]<-2$. Stars from the SAGA database are selected with iron abundance in the range $[-2.8,\,-1.8]$ to increase the number statistics. In all cases CEMP stars are defined as stars with carbon abundance $[\C/\Fe]>+1$.
}
\label{tab:cf_A13}
\centering
\begin{tabular}{ c  c }
\hline
\hline
	model sets and data &	$\fC~(\%)$		\\
\hline
A1										&	$8.05\pm0.02$	\\
A4										&	$7.91\pm0.02$	\\
C$_{\mathrm{q}}$ \cite[][]{Abate2013}	&	$2.38\pm0.04$	\\
\hline
SEGUE								&	$6.1\pm1.0$	\\
SAGA database							&	$23.5\pm2.9$	\\

\hline
\end{tabular}
\end{table}


\begin{figure}[!t]
\centering
\includegraphics[width=0.48\textwidth]{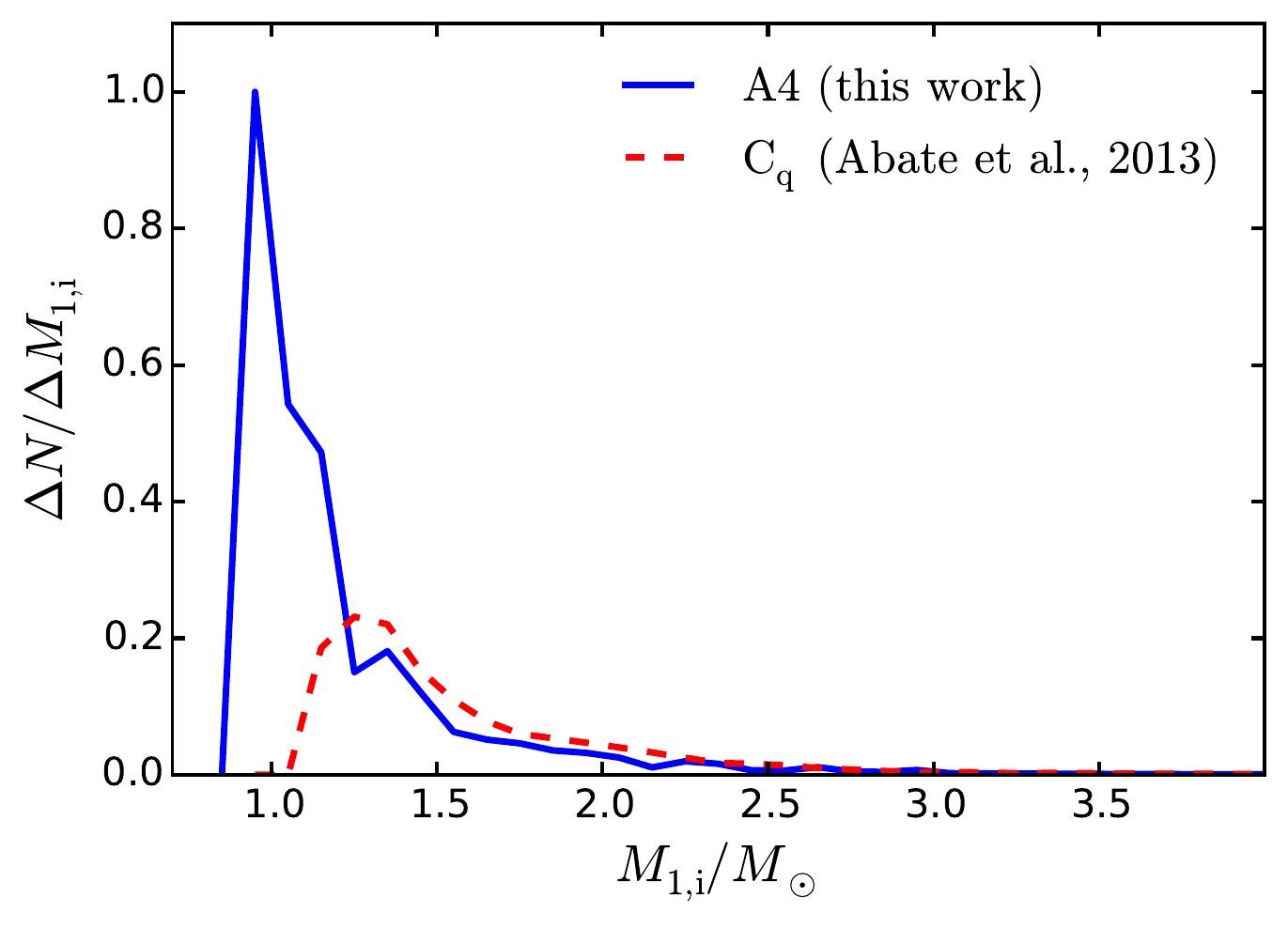}
\caption{Distribution of $\Mprim$ computed with model set A4 (solid, blue line) and model set $\mathrm{C_q}$ of \cite{Abate2013} (dashed, red line). 
The bins are equally spaced and the width in $\Mprim/\Msun$ of each bin is $0.1$. 
The $y$-axis indicates the number of CEMP stars in each bin, arbitrarily rescaled.
}
\label{fig:M1i_A}
\end{figure}

With these assumptions, the only difference between model sets A4 and
$\mathrm{C_q}$ is in the adopted set of AGB evolution models. The AGB
model of \cite{Abate2013} was tuned to reproduce the evolution of AGB
stars as predicted by the detailed models of \cite{Karakas2002} and
\cite{Karakas2007}. In these models, AGB stars of mass below
approximately $1.2\,\Msun$ did not experience third dredge-up, and
consequently did not contribute to the formation of CEMP stars. On the
other hand, \cite{Stancliffe2008} and the recent work of
\cite{Karakas2010}, on which our AGB model is based, show that AGB stars
of mass down to $0.9\,\Msun$ undergo third dredge-up. Consequently, in
our model set A4 CEMP stars are formed in a wider range of $\Mprim$. The
distributions of $\Mprim$ computed with model sets A4 and $\Cq$ are
compared in Fig. \ref{fig:M1i_A}. Model set A4 peaks for $\Mprim$ in the
range $[0.9,\,1.1]\,\Msun$, in which set $\Cq$ does not produce CEMP
stars. With the adopted solar-neighbourhood IMF low-mass stars are much
more commo, consequently model set A4 predicts a CEMP fraction of
$7.9\%$, i.e. a factor $3.3$ larger than set $\Cq$, as shown in
Table~\ref{tab:cf_A13}. 

Model set A1 differs from model set A4 in that the model stars are
selected according to their luminosity, as described in Sect.
\ref{method}. With model set A1 we find a CEMP/VMP ratio of $8.0\%$. 
Comparison with the result of model set A4 indicates that the selection
criterion has a small impact on the total fraction of CEMP stars because
in our synthetic population the fraction of
CEMP stars is very similar among stars above and below the threshold of
$\loggunits=4$. If we reduce the minimum abundance of carbon in the
definition of CEMP stars, and adopt $[\C/\Fe]>+0.7$ (our
default assumption in this paper), the CEMP/VMP ratio increases to
$8.5\%$ (model set A, Table \ref{tab:SEGUE}). This occurs because
our model produces very few stars with $+0.7<[\C/\Fe]\le+1.0$,
whereas $94\%$ of the model CEMP stars have $[\C/\Fe]>+1$, and the
average carbon abundance in the synthetic CEMP population is
$[\C/\Fe]=+1.9$.

Table \ref{tab:cf_A13} compares the model predictions with the observed
CEMP/VMP ratio derived from the SEGUE and SAGA data sets. The CEMP
fraction among SEGUE stars is taken from Table 4 of \cite{Lee2013} for
stars with iron abundance in the range $-2.5<[\Fe/\Hy]\le-2$ and with
$[\C/\Fe]>+1$. A more detailed comparison between the results of our
model sets and the SEGUE sample is performed in the next section. The
CEMP fraction in the SAGA sample is approximately $24\%$, almost four
times larger than the value determined for SEGUE stars. We discuss this
discrepancy in Sect. \ref{discussion}.

\subsection{The fraction of CEMP stars}

We now compare the CEMP fraction observed in the SEGUE sample with the
results of our simulations. We select stars according to the method
explained in Sect. \ref{method}, with $\Vmin=13.5$,
$\Vmax=18.5$ and a carbon abundance in CEMP stars $[\C/\Fe]>+0.7$. 

Table \ref{tab:SEGUE} summarises the results obtained with
different model sets. The observed fractions and relative uncertainties
are computed from the data reported by \citet[Table 6]{Lee2013} for
stars in the range $-2.5\le[\Fe/\Hy]\le-2$. For consistency with the
assumptions of \cite{Lee2013}, a minimum observational uncertainty of
$1\%$ is adopted, and stars are defined as giant, turnoff, and dwarf if
their surface gravities are $\loggunits<3.5$, $3.5\le\loggunits<4.2$ and
$\loggunits\ge4.2$, respectively. The CEMP/VMP ratios computed among
giants, turnoff, and dwarf stars are indicated as $\fCgia$, $\fCTO$,
$\fCdwa$ (columns $3-5$). The proportions of giant, turnoff and dwarf
stars among CEMP stars are indicated as $\fgiaC$, $\fTOC$, $\fdwaC$,
respectively (columns $6-8$). The quoted uncertainties in the models
convey only Poisson statistics. In this section we compare the
observed and model fractions calculated with our default model set A,
while the comparison of the results with different model sets is the
subject of the next section.

\begin{figure}[!t]
\centering
\includegraphics[width=0.48\textwidth]{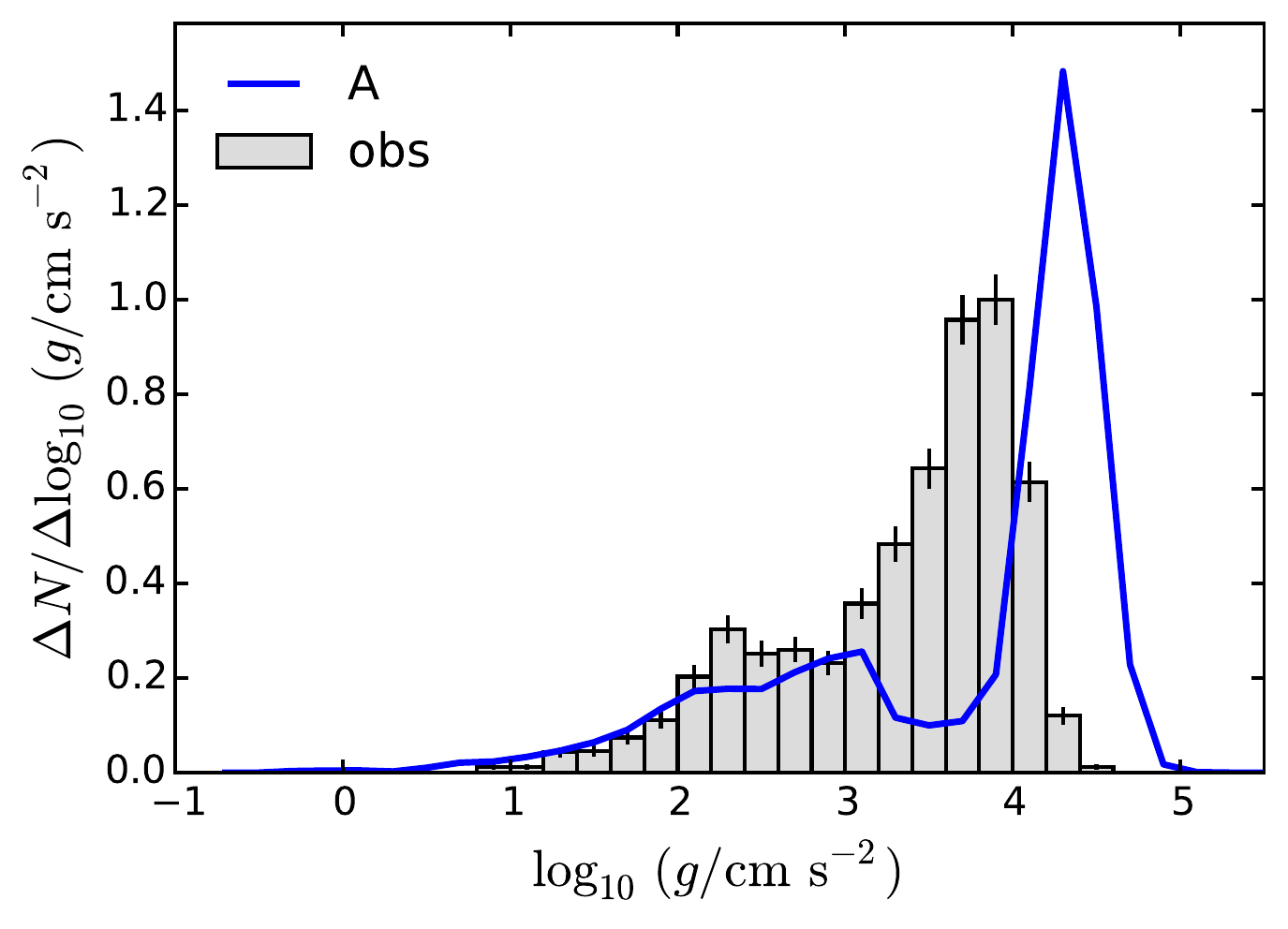}
\caption{Distribution of $\loggunits$ as derived from the SEGUE sample of CEMP stars with $-2.5\le[\Fe/\Hy]\le-2$ (histogram with Poisson errors) and computed with our default model set A (solid line). The $0.2$ dex bins are equally spaced in $\loggunits$. In this and the similar plots that follow the area under the graph (which is the total number of stars) is normalised such that it is the same for both the observations and our model stars.}
\label{fig:logg19}
\end{figure}

Our total CEMP/VMP fractions underestimate those observed by a factor of $1.2-1.4$.
We note a large difference between the observed and modelled proportions of the three
evolutionary stages among CEMP stars. In our models almost half of the
CEMP stars are on the main sequence. By contrast, in the observed sample
the CEMP dwarf stars count for approximately $1.6\%$, while giants and
turnoff stars count for approximately $50\%$ each. This discrepancy is
also evident from the comparison between the $\logg$ distribution of the
observed CEMP stars of the SEGUE sample and the results of model set A
(Fig. \ref{fig:logg19}). The positions of the peaks of the two
distributions differ by approximately $0.5$ dex. If we artificially
decrease the $\logg$ of our model CEMP stars by $0.5$ dex to mimic a
hypothetical systematic offset we obtain fractions of giant, turnoff, and
dwarf stars among CEMP stars of $35.4\%$, $60.5\%$ and $4.1\%$,
respectively. However, with such a systematic offset we find that
approximately $3\%$ of the modelled CEMP stars have $\loggunits<1$,
while far fewer SEGUE stars are observed at such low gravity \cite[][]{Lee2013}.

\begin{figure}[!t]
\centering
\includegraphics[width=0.5\textwidth]{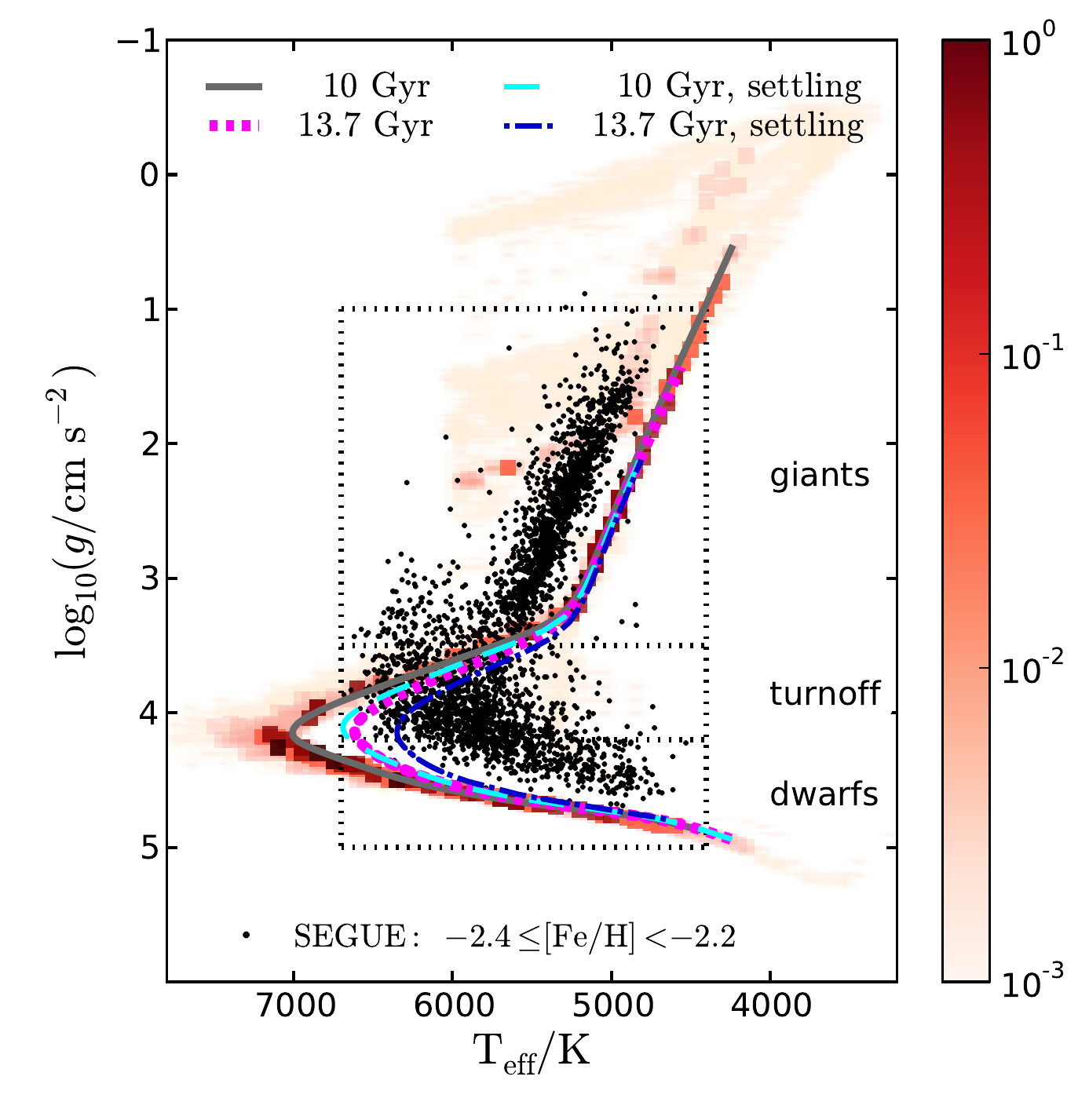}
\caption{$\Teff-\logg$ probability distribution of all stars in our simulation computed with model set A. Darker colours indicate regions of higher probability. The grey solid and magenta dotted lines show the isochrones computed with detailed stellar-evolution models at $10$ Gyr and $13.7$ Gyr, respectively. The isochrones computed for the same ages with a detailed model that includes gravitational settling are shown as light-blue dashed and blue dot-dashed lines, respectively. The black dotted lines indicate the ranges of $\Teff$ and $\logg$ in which \cite{Lee2013} select the SEGUE stars. Black circles indicate all SEGUE stars with $-2.4\le[\Fe/\Hy]<-2.2$.}
\label{fig:loggT_19}
\end{figure}

An offset in the model-vs.-observed surface gravities can also be
identified in Fig. \ref{fig:loggT_19}, in which the temperature-gravity
distribution of our synthetic population (red-scaled distribution) is compared
to SEGUE stars with iron abundances in the range
$-2.4\le[\Fe/\Hy]<-2.2$. The surface gravity of the main sequence and of
the turnoff point in the synthetic population appear to be shifted by
several times $0.1$ dex with respect to the observed sample. The reason
for such a discrepancy is currently unclear. One possible explanation is
that in our simulations we overestimate the upper limit in $V$~magnitude
and, as a consequence, in our selection procedure we take into account
stars that in reality are too faint to be detected. However, Table
\ref{tab:SAGA} shows that, even if $V_{\mathrm{max}}$ is reduced by 2
mag, the proportion of dwarfs among CEMP stars decreases only by a
factor of two. To reproduce the observed dwarf fraction a far lower
$V_{\mathrm{max}}$ is necessary, which is inconsistent with the
characteristics of the SEGUE survey. An alternative explanation of this
discrepancy is that the surface gravities determined from the SEGUE
spectra are systematically underestimated. However, the algorithms to
determine the surface parameters of SEGUE stars are calibrated on a set
of stars with available high-resolution spectroscopy, and systematic
errors that shift $\logg$ by more than $-0.3$~dex are considered highly
unlikely \cite[][]{Lee2013}.

%
%

Figure \ref{fig:loggT_19} shows that there is also a discrepancy between
the effective temperatures of the modelled and observed stars. To
reliably estimate [C/Fe], \cite{Lee2013} restrict the SEGUE
sample to stars with $4,\!400\,\mathrm{K}\le\Teff\le6,\!700\,
\mathrm{K}$ because in this range their technique is robust. However, in our
models most of the turnoff stars have $\Teff>6,\!700$ K, up to
$\approx8,\!000$ K, regardless of the predicted abundance of carbon,
whereas in the SEGUE sample there are very few turnoff stars with
$\Teff>6,\!700$~K (which are not plotted in Fig. \ref{fig:loggT_19}).
Such a discrepancy is at least partly explained by an inaccuracy in our
model temperatures. Our code calculates the effective temperature from
the luminosity and the radius of the star, which are computed with
fitting formulae based on the detailed stellar models of
\cite{Pols1998}. For a $0.8\,\Msun$ star at the turnoff point these
fitting formulae reproduce the luminosity and radius of the detailed
model within a factor of $1.15$ and $0.98$, respectively. However,
because with $\binc$ the luminosity is higher and the radius is smaller
than in the detailed model, the combination of these two factors gives a
higher effective temperature than in the detailed model.
Fig.~\ref{fig:loggT_19} shows two isochrones, for $10\,$Gyr and $13.7\,
$Gyr (grey solid and magenta dotted lines), which were constructed using
the same version of the STARS code (\citealp{Eggleton1971}, subsequently
updated by many authors, e.g. \citealp{Pols1995}) as adopted in
the detailed models of \cite{Pols1998}. These isochrones bracket the age
range of our synthetic population. The offset between our
population-synthesis results and these isochrones yields an estimate of
the error in temperature because of inaccuracies in the fitting
formulae.

\begin{table*}[!t]
\caption{CEMP star fraction computed from the SEGUE data in the range $-2.5<[\Fe/\Hy]\le2$, and computed with our different model sets. The range of $V$ magnitudes adopted in the models is $[13.5,\,18.5]$. $\fC$, $\fCgia$, $\fCTO$, $\fCdwa$ indicate, respectively, the CEMP/VMP ratio computed among all/giant/turnoff/dwarf stars. The fraction of giants, turnoff stars and dwarf stars computed among CEMP stars is indicated respectively as $\fgiaC$, $\fTOC$, $\fdwaC$ (the latter fractions add up to $100\%$). The errors in the models convey only Poisson statistics. A minimum observational uncertainty of $1\%$ is adopted \cite[cf.][]{Lee2013}. CEMP stars have carbon abundance $[\C/\Fe]>+0.7$.
}
\label{tab:SEGUE}
\centering
\begin{tabular}{c   r   r r r   r r r}
\hline
	model sets and data	&	$\fC\,(\%)~~~$	&	$\fCgia\,(\%)~~$	&	$\fCTO\,(\%)~~$	&	$\fCdwa\,(\%)~~$	&	$\fgiaC\,(\%)$	&	$\fTOC\,(\%)$	&	$\fdwaC\,(\%)$	\\
\hline\\[-1ex]
SEGUE		&	$11.46\pm1.00$		&	$15.79\pm1.00$		&	$10.38\pm1.00$		&	$1.67\pm1.00$	&	$51.42 \pm 2.30$	&	$47.02 \pm 2.16$	&	$1.56 \pm 1.00$		\\[1ex]
\hline\\[-1ex]
A			&	$8.50 \pm 0.02$		&	$9.59 \pm 0.03$		&	$10.73 \pm 0.04$	&	$7.32 \pm 0.02$	&	$31.95 \pm 0.10$	&	$19.80 \pm 0.08$	&	$48.25 \pm 0.13$	\\
B			&	$8.96 \pm 0.02$	&	$10.30 \pm 0.03$	&	$11.01 \pm 0.04$	&	$7.72 \pm 0.02$	&	$32.47 \pm 0.10$	&	$19.23 \pm 0.07$	&	$48.30 \pm 0.13$	\\
C			&	$9.48 \pm 0.02$	&	$10.67 \pm 0.03$	&	$11.78 \pm 0.04$	&	$8.24 \pm 0.02$	&	$31.69 \pm 0.10$	&	$19.56 \pm 0.07$	&	$48.75 \pm 0.13$	\\
D			&	$12.17 \pm 0.02$	&	$9.94 \pm 0.03$	&	$15.21 \pm 0.05$	&	$12.45 \pm 0.02$		&	$23.12 \pm 0.07$	&	$19.60 \pm 0.06$	&	$57.28 \pm 0.12$	\\[1.2ex]

$\IK$		&	$8.81 \pm 0.01$	&	$9.98 \pm 0.03$	&	$11.12 \pm 0.04$	&	$7.58 \pm 0.02$			&	$31.99 \pm 0.10$	&	$19.75 \pm 0.08$	&	$48.27 \pm 0.13$	\\
$\IL$	&	$16.47 \pm 0.02$	&	$17.23 \pm 0.04$	&	$18.31 \pm 0.05$	&	$15.40 \pm 0.03$	&	$32.23 \pm 0.08$	&	$19.27 \pm 0.06$	&	$48.50 \pm 0.10$	\\[1.2ex]

Qp4		&	$7.87 \pm 0.01$		&	$8.69 \pm 0.03$	&	$9.85 \pm 0.04$	&	$6.90 \pm 0.02$				&	$31.37 \pm 0.11$	&	$19.45 \pm 0.08$	&	$49.18 \pm 0.14$	\\
Q1		&	$11.37 \pm 0.02$	&	$12.18 \pm 0.03$	&	$13.95 \pm 0.05$	&	$10.11 \pm 0.02$	&	$32.27 \pm 0.10$	&	$20.28 \pm 0.07$	&	$47.45 \pm 0.13$	\\[1.2ex]

$\SDM$	&	$11.16 \pm 0.02$	&	$12.45 \pm 0.03$	&	$14.06 \pm 0.04$	&	$9.67 \pm 0.02$		&	$31.98 \pm 0.08$	&	$19.72 \pm 0.06$	&	$48.29 \pm 0.11$	\\
S3		&	$9.41 \pm 0.02$		&	$10.72 \pm 0.03$	&	$11.82 \pm 0.04$	&	$8.08 \pm 0.02$		&	$31.71 \pm 0.09$	&	$20.05 \pm 0.07$	&	$48.24 \pm 0.12$	\\[1.2ex]

T8		&	$9.89 \pm 0.01$		&	$12.14 \pm 0.03$	&	$10.11 \pm 0.03$	&	$8.78 \pm 0.02$		&	$30.91 \pm 0.09$	&	$20.22 \pm 0.07$	&	$48.86 \pm 0.12$	\\
T12		&	$8.14 \pm 0.02$		&	$9.72 \pm 0.04$	&	$13.60 \pm 0.08$	&	$6.43 \pm 0.02$			&	$31.52 \pm 0.16$	&	$19.70 \pm 0.12$	&	$48.78 \pm 0.21$	\\
\hline
\end{tabular}
\end{table*}

The discrepancy between modelled and observed temperatures decreases if
efficient gravitational settling is included in the models. Two
isochrones including settling are shown in Fig.~\ref{fig:loggT_19}
(light-blue dashed and blue dot-dashed lines, for $10\,$Gyr and $13.7\,
$Gyr, respectively), which were constructed using a recent version of
the STARS code \cite[][]{Stancliffe2008,Stancliffe2009-3}.
Settling decreases the turnoff
temperature of the $10\,$Gyr isochrone by approximately $300$~K, that is,
the same shift caused by a $3\,$Gyr difference in age. The $13.7\,$Gyr
isochrone with settling has a turnoff temperature of about $6400$~K, which
is not much hotter than the average turnoff temperature of the observed
stars ($\approx6000$~K). Although the effects of gravitational settling
on the structure and composition of metal-poor stars are not fully
explored \cite[][]{Richard2002-3, Richard2002-1, Stancliffe2008}, these
results suggest that the discrepancy in effective temperature can be
reduced by including settling in our models.

The effective temperature of giant stars is underestimated both by the
population synthesis models and by the isochrones. This discrepancy can also
be resolved by calibrating the mixing-length parameter in the detailed
models \cite[as done for example by][in the Yonsei-Yale isochrone
sets]{Spada2013}, and it does not affect any of the results presented in
this work.

In the observed sample the CEMP/VMP ratio is found to increase
significantly with luminosity, that is, $\fC$ is smaller among dwarf
stars and larger among giants, although this relation is reversed at
lower metallicity ($\Fe/\Hy<-3$, \citealp{Lee2013}). In our models we
find smaller variations of the CEMP/VMP ratio among the different
classes of stars. The smallest CEMP/VMP ratio is calculated among dwarf
stars and the largest among turnoff stars. Turnoff and giant CEMP stars
in the models have very similar mass distributions: the CEMP fraction is
smaller among the giants as an effect of the first dredge-up, which
depletes part of the carbon at the surface. Consequently, a turnoff CEMP
star with carbon abundance close to the threshold $[\C/\Fe]=+0.7$ has
$[\C/\Fe]<+0.7$ after the dredge-up. This effect is small because in our
model thermohaline mixing already dilutes the accreted material
throughout the entire star. If we inhibit thermohaline mixing in our
models the difference between $\fCgia$ and $\fCTO$ increases (Table
\ref{tab:SEGUE}, model set D).

The reason for the discrepancy between the predictions of our models and
the observations is still unclear. One possible explanation is related
to the method adopted by \cite{Lee2013} to compute $\fC$. Stars for
which only an upper limit to [C/Fe] is available or stars with a poor
carbon measurement (i.e. if the correlation coefficient between the
observed and synthetic spectrum is lower than $0.7$) are counted among
carbon-normal very metal-poor stars. Carbon lines are more difficult to
detect in the spectra of warm stars, which are therefore clearly
identified as CEMP stars only if carbon is very enhanced. As a
consequence, the CEMP/VMP ratio computed for turnoff stars is somewhat
biased towards lower values, although this effect is probably small
\cite[][]{Lee2014}. This bias is not expected to be significant for the
observed dwarf stars, which are typically cooler than turnoff stars. The
fact that the observed $\fCdwa$ is approximately a factor of ten lower
than $\fCTO$ and $\fCgia$ suggests some effect prevents the detection of
dwarf CEMP stars. It is still unclear whether this is a physical effect,
which should be stronger for carbon-rich dwarfs than for carbon-normal
ones, or a bias introduced during the spectral analysis.

\subsection{CEMP fractions computed adopting different sets of initial parameters}

The characteristics of our synthetic CEMP populations depend on our
assumptions about the mass-transfer process, the extent to which the
accreted material is diluted, and the properties of the stellar
population, such as its age, and the initial distributions of masses and
separations.

\begin{figure}[!t]
\centering
\includegraphics[width=0.48\textwidth]{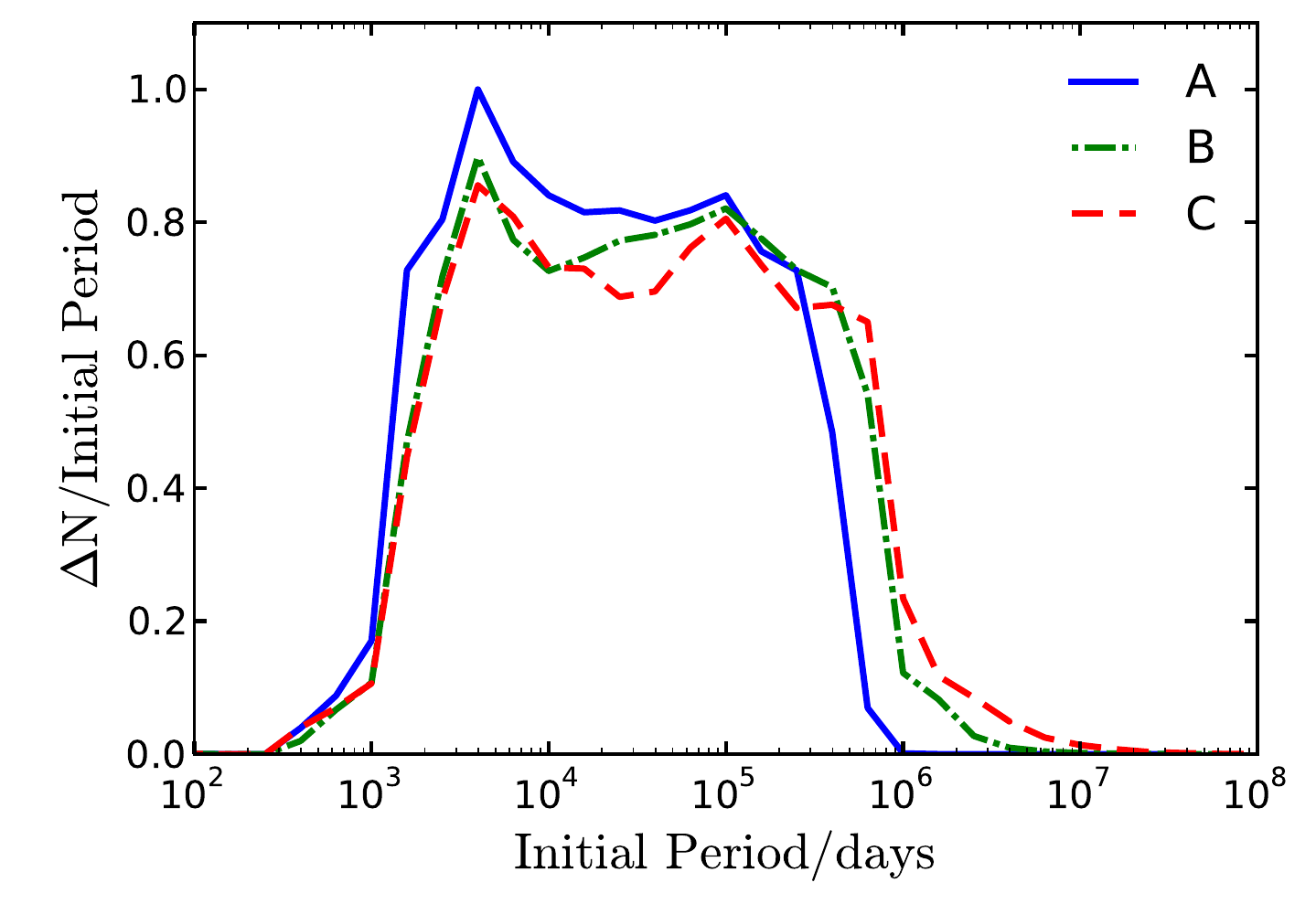}
\caption{Initial period distribution computed with model sets A, B, and C (solid, dot-dashed and dashed line, respectively). The $0.2$ dex bins are equally spaced in $\log_{10}(\Pin/\mathrm{days})$.}
\label{fig:peri}
\end{figure}

The differences in the CEMP/VMP ratios between model sets A, B, and C are
mainly due to the range of periods at which the secondary star accretes
material from the AGB primary star, as shown in Fig. \ref{fig:peri}.
Model sets B and C have a larger fraction of CEMP stars than
set A because of the more efficient mechanism of angular-momentum loss
which allows binary systems to interact at longer initial periods.
The largest range of initial periods is accessible to model set C
because it combines the model of efficient angular-momentum loss with
the WRLOF model of wind accretion. Similar results are found by
\cite{Abate2013}. In binary systems that are initially very wide
($\Pin>10^6$ days) only relatively massive primary stars ($\Mprim>1.5\,
\Msun$) transfer significant amounts of mass. Such stars are rare
according to the solar-neighbourhood IMF adopted in our models.
Consequently the increase in $\fC$ in models B and C is rather
modest.

The relative proportion of giant, turnoff, and dwarf stars depends
essentially on the mass distribution of the CEMP stars. All model sets
produce the majority of CEMP stars with masses after accretion between
$0.75\,\Msun$ and $0.9\,\Msun$. However, the initial mass distribution
of the CEMP stars and the distribution of the accreted masses
differ between model sets. In wide binary systems ($P>10^4$ days) our
WRLOF model typically predicts larger accretion efficiencies than our
enhanced Bondi-Hoyle-Lyttleton model. Hence, at large separations
secondary stars on average accrete more mass with model sets A and C
than does model set B, and are initially less massive. As a
consequence, the proportion of CEMP stars formed from systems with
initial secondary masses less than $0.7\,\Msun$ is $41\%$ with model
sets A and C, while it is $24\%$ with model set B.

In our model set D we inhibit the thermohaline mixing, consequently
the material accreted by the secondary stars remains on the surface
until it is mixed in by convection. As a result, the range of initial
periods in which CEMP stars are formed increases (by roughly a factor of $25$) 
because even a tiny amount of mass transferred from the AGB primary star on top of a dwarf or
turnoff star is sufficient to enhance carbon. Consequently, the
fractions $\fCTO$ and $\fCdwa$ both increase, as does the overall
CEMP/VMP ratio.

Our knowledge of the initial properties of the halo stellar population
is very uncertain and difficult to constrain. To determine how robust
our results are to changes in the age, the IMF, and the initial
distributions of separations and mass ratios of our models, we simulate
populations of VMP binary stars with a variety of model sets that adopt
different assumptions. The CEMP/VMP ratio typically varies between $7\%$
and $12\%$, with the exception of model $\IL$, while the proportion of
different stellar types among CEMP stars is roughly constant. Below we briefly
summarise the results obtained with the different model sets.

\begin{figure*}[!t]
\includegraphics[width=0.498\textwidth]{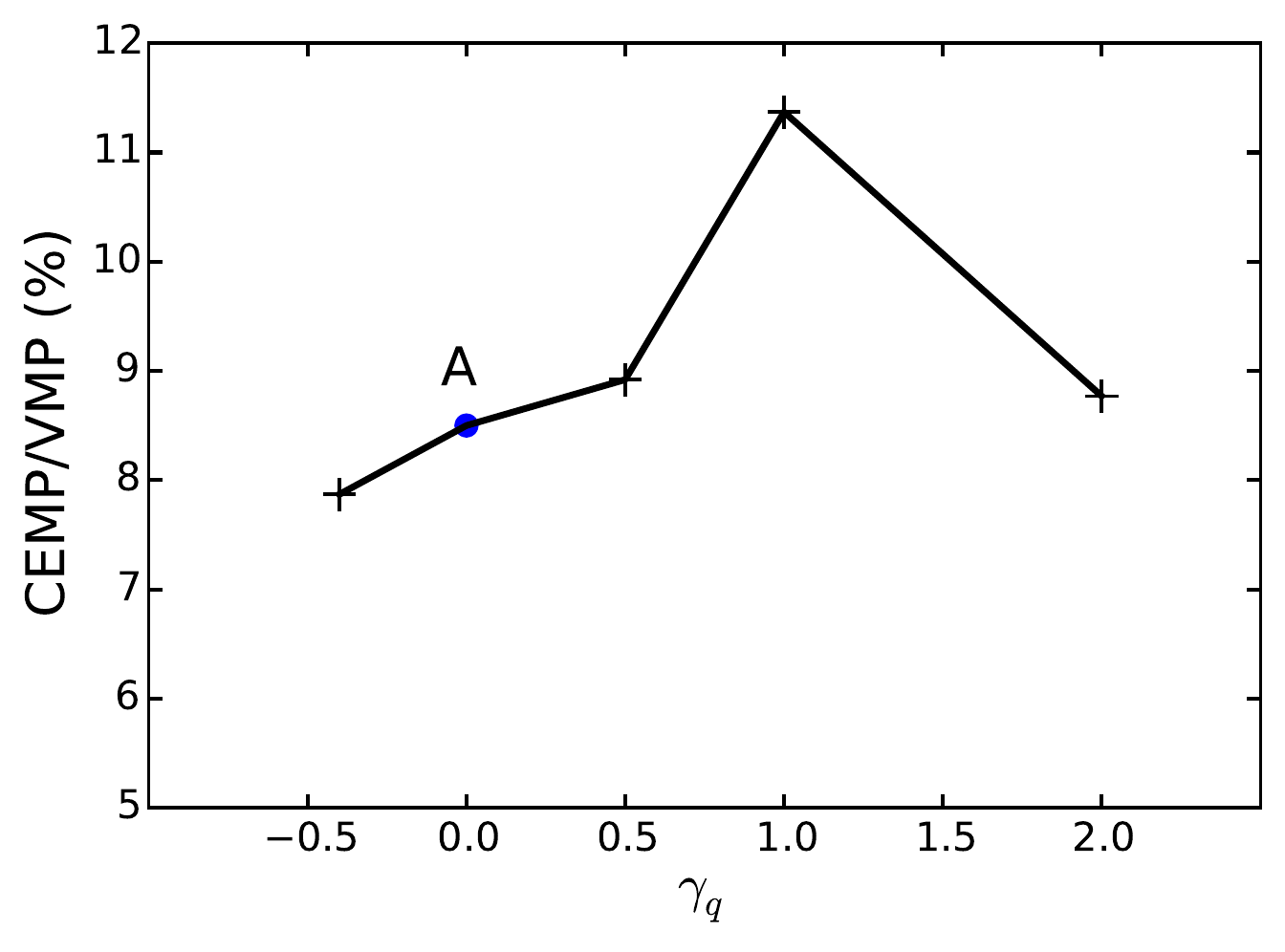}
\includegraphics[width=0.498\textwidth]{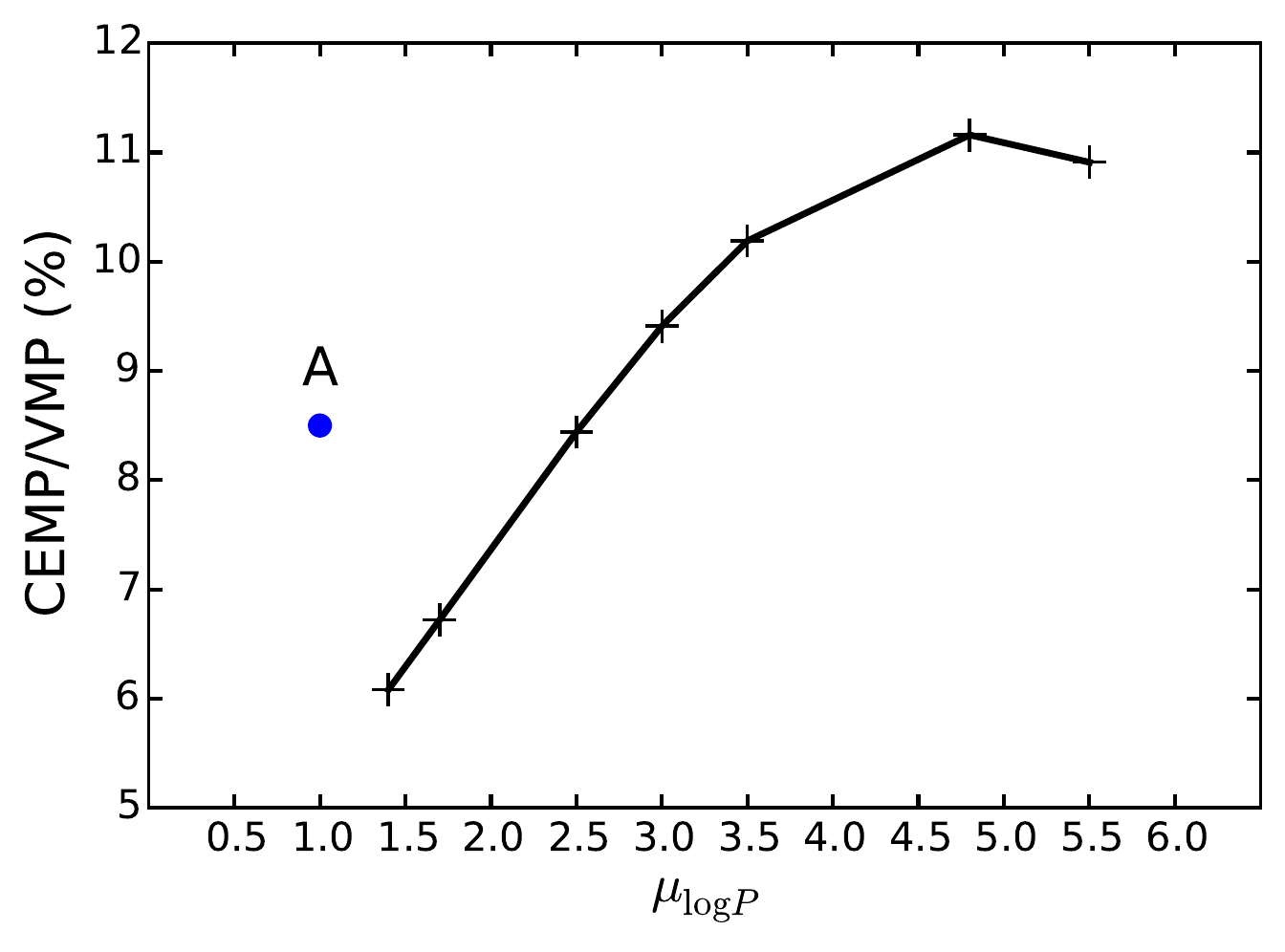}
\caption{{\it Left:} $\fC$ calculated with model sets that assume a power-law initial mass-ratio distribution, $\phi(q)\propto q^{\gamma_q}$, with different values of $\gamma_q$.
{\it Right:} $\fC$ of model sets that assume a log-normal initial-period distribution with different values of the mean, $\mu_{\log P}$, and fixed standard deviation, $\sigma_{\log P}=2.3$. The filled circles show the value of $\fC$ computed with our default model set A.
}
\label{fig:q-sep-models}
\end{figure*}

\begin{itemize}

\item[$\bullet$] {\it Variations in the IMF (model sets $I_{\mathrm{K01}}$ and $I_{\mathrm{L05}}$)}.
With the IMF proposed by \cite{Kroupa2001} primary stars above one solar
mass are favoured compared to the default IMF because the IMF slope is
slightly shallower; therefore, with model set $\IK$ there is a small
increase in $\fC$. Compared to the default IMF, the IMF suggested by
\cite{Lucatello2005a} produces more stars with $M\ge0.9\,\Msun$ (which
can contribute to the formation of CEMP stars) than stars with $M<0.9\,
\Msun$ (which do not form CEMP stars), therefore $\IL$ is the model that
predicts the largest CEMP/VMP ratio.
\item[$\bullet$] {\it Variations in the initial mass-ratio distribution 
(model sets~Qp4 and Q1)}. Model set Qp4, with $\phi(q)\propto q^{-0.4}$,
favours systems with low mass ratios, $q=\Msec/\Mprim$. Binary systems
with low-mass secondary stars ($\Msec<0.5\,\Msun$) do not contribute
significantly to the CEMP fraction, either because the secondary stars
need to accrete large amounts of mass, or they are not sufficiently
luminous to be detected. As a consequence, $\fC$ decreases. On the
contrary, model set Q1, with $\phi(q)\propto q$, favours equal mass
ratios; therefore, binary systems with relatively massive secondary
stars are more common, which need to accrete a small amount of mass to
become sufficiently luminous CEMP stars. Consequently, $\fC$ increases.
In the left panel of Fig. \ref{fig:q-sep-models} we show the value of
$\fC$ computed with five model sets that assume different
$\gamma_q$ in a power-law distribution of the initial
mass-ratio, $\phi(q)\propto q^{\gamma_q}$. The maximum CEMP/VMP ratio is
found with $\gamma_q=1$, i.e. with model set Q1.
\item[$\bullet$] {\it Variations in the initial-period distribution (model sets $S_{\mathrm{\!DM}}$ and S3).}
Because the initial-period distribution of halo binary stars is poorly
constrained, we test a variety of models with log-normal period
distributions in which we vary the mean, $\mu_{\log P}$, while we adopt
a fixed standard deviation, $\sigma_{\log P}=2.3$, i.e. the value
determined by \cite{DuquennoyMayor1991} for a solar-neighbourhood
population of binaries with G-dwarf primaries. The CEMP/VMP ratio
computed with these models is shown in the right panel of
Fig.~\ref{fig:q-sep-models} as a function of $\mu_{\log P}$. The maximum
is found with model set $\SDM$, that is, the \cite{DuquennoyMayor1991}
distribution with $\mu_{\log P}=4.8$. With model set $\SDM$, the initial
period distribution peaks at orbital periods between $5\times10^4$ and
$10^5$ days, which is approximately the range where our WRLOF model of
wind accretion is most efficient in forming CEMP stars. Because the
average orbital period in the sample of $15$ observed CEMP-$s$ stars
analysed in Paper I is approximately $1,\!500$ days, we assume
$\mu_{\log P}=3$ in our model set S3. The CEMP/VMP ratio decreases
compared to model $\SDM$ because model S3 predicts more close binary
systems that undergo a common-envelope phase and do not form CEMP stars.
On the other hand, model set S3 predicts a larger CEMP/VMP ratio than
our default model set A (with a flat $\log P$-distribution) because it
has fewer very wide systems that do not interact. If a log-normal
period distribution (as in model $\SDM$) is combined with a mass-ratio
distribution that favours equal masses (as in model Q1), a CEMP/VMP
fraction of approximately $15\%$ is produced, i.e. almost the same
fraction as assuming an IMF weighted towards relatively large masses,
$M\gtrsim0.9\Msun$.
\item[$\bullet$] Variations in the minimum age of the synthetic population (model sets T8 and T12).
A change in the minimum age of the stellar population in the halo by
$\pm2$ Gyr has a small effect on $\fC$, as also found by \cite{Izzard2009}. This is mainly because of the
difference in lifetime between the stars that accreted mass during the
evolution and the stars that did not. As we discuss in more detail in
Sect. \ref{mt}, for a minimum age of the halo of $t_{\mathrm{min}}=10\,
$Gyr, stars of mass $0.9\Msun$ and above have already become white
dwarfs, whereas CEMP stars that (after accretion) have masses up to about
$0.95\Msun$ may still be visible. That is, only CEMP stars can be
detected in the mass range $[0.9,0.95]\Msun$ for this age
(Fig.~\ref{fig:mt}). With $t_{\mathrm{min}}=8\,$Gyr, CEMP stars up to
about $1.05\Msun$ are still visible, whereas normally at this age stars
more massive than $0.95\Msun$ have already become white dwarfs, that is,
the difference in the maximum mass of visible stars increases. As a
result, the CEMP fraction increases. On the other hand, with
$t_{\mathrm{min}}=12\,$Gyr such a difference in mass decreases (the
maximum masses are approximately $0.88\Msun$ and $0.91\Msun$ for normal
stars and CEMP stars, respectively), and so does the CEMP fraction.
\end{itemize}

\subsection{Orbital periods of the CEMP stars.}
\label{Pf}

\begin{figure}[!t]
\centering
\includegraphics[width=0.48\textwidth]{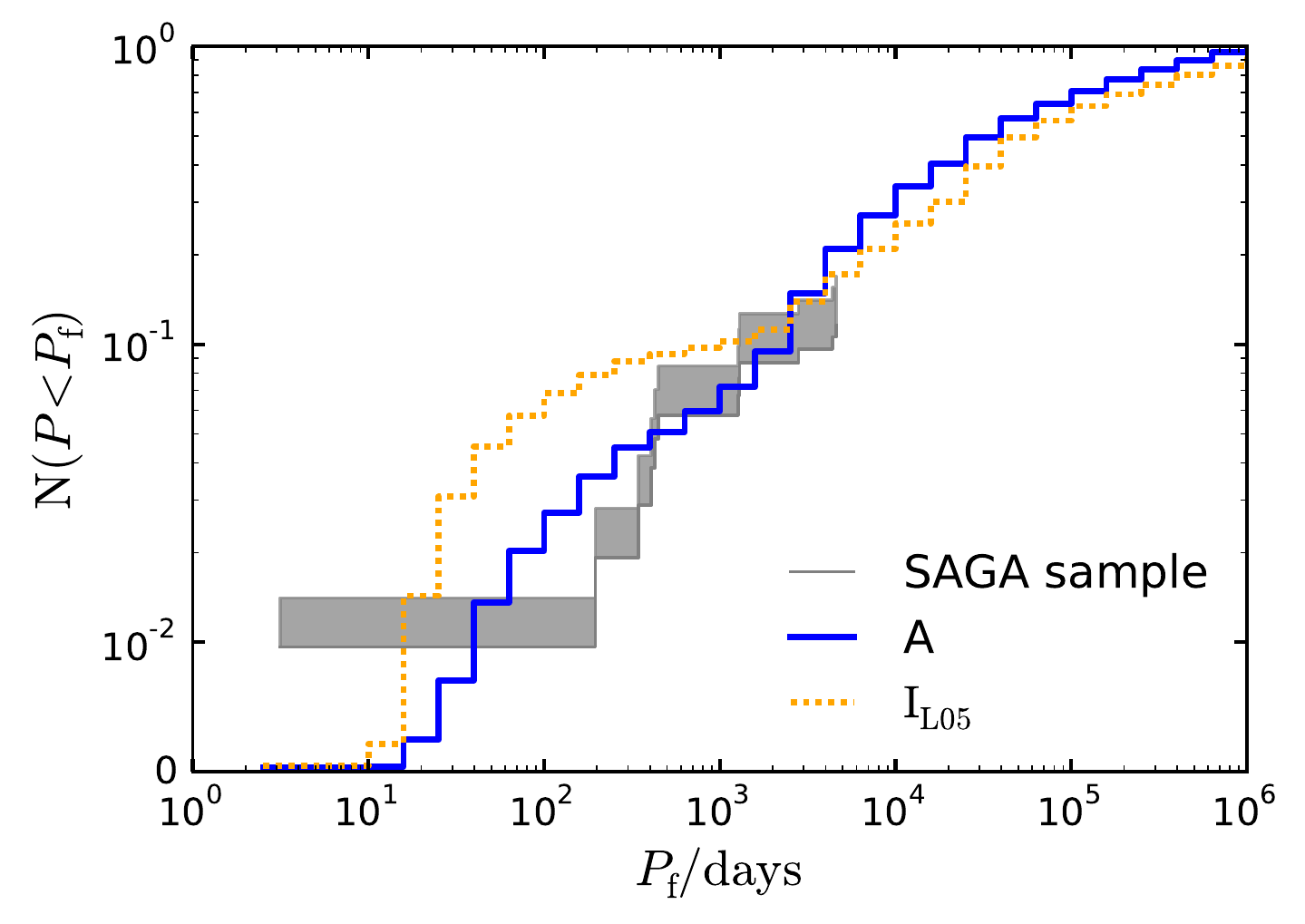}
\caption{Cumulative orbital-period ($\Pf$) distributions for SAGA CEMP stars with $-2.8\le[\Fe/\Hy]\le-1.8$ (shaded area). The lower limit of the observed distribution is determined assuming that all CEMP stars are binaries, whereas to determine the upper limit we assume that only CEMP-$s$ stars are in binaries. Our model sets A and $\IL$ are shown as solid and dotted lines, respectively. The $0.2$ dex bins of the model distributions are equally spaced in $\log_{10}(\Pf/$days).}
\label{fig:cumul_Pf}
\end{figure}

We now compare the orbital-period distribution predicted by our models with
the observed sample of $15$ binary CEMP stars with measured periods
studied in Paper I. We exclude from this sample the stars CS~22964--161
A and B because they have most likely been polluted by a third
star whose period is unknown \cite[][]{Thompson2008}, LP~625--55 because
it only has a lower limit on the orbital period, and CS~29497--034
which has $[\Fe/\Hy]=-2.96$. Our models produce most CEMP stars in
orbits longer than a few thousand days, whereas the observed systems
have orbital periods shorter than $4,\!600$ days. However, in the SAGA sample,
only $11$ out of $94$ CEMP stars with $-2.8\le[\Fe/\Hy]\le-1.8$ have
measured orbital periods, thus they may not be representative of the
period distribution of the entire sample. To compare the observations
with the results of our models, Fig. \ref{fig:cumul_Pf} shows the
observed and modelled cumulative period distributions. The implicit
assumptions made in deriving the observed cumulative distribution are
($i$) that all CEMP stars are formed in binaries, most of which have
unknown periods, and ($ii$) that the SAGA sample is complete for
$\Porb\le4,\!600$ days. The first assumption is probably correct for
CEMP-$s$ stars, which are mostly found in binary systems, but it is
unlikely to be valid for CEMP-no stars (\citealp{Lucatello2006};
\citealp{Norris2013-2}; \citealp{Starkenburg2014}; Hansen et al. {\it in prep.}). 
For this reason we plot the cumulative
period distribution as a range (shaded area in Fig. \ref{fig:cumul_Pf}),
in which the lower limit corresponds to the hypothesis that all $94$
CEMP stars in the SAGA sample are binaries, while the upper limit
corresponds to the hypothesis that only the $71$ CEMP-$s$ stars in the
sample are binaries. The solid and dotted lines in Fig.
\ref{fig:cumul_Pf} represent the cumulative period distributions
predicted with models A and $\IL$, respectively, which bracket the
distributions found with all the other models. 

Under the above assumptions, our models are consistent with the
observations in the range between approximately $600$ and $5,\!000$
days, and they seemingly predict an excess of CEMP stars with periods
between a few tens and a few hundreds of days. On the other hand, the
models do not reproduce the single observed CEMP star%
\footnote{Possible formation scenarios for this peculiar CEMP star
with period $3.14$ days, HE~0024--2523, are discussed, for example, by
\cite{Lucatello2003} and in Paper I.} %
at $\Pf < 10$ days. However,
the assumption that the unknown periods of all other CEMP stars in our
sample are longer than about $5,\!000$ days may be incorrect; in reality
at least some of these stars are likely to have shorter periods
\cite[e.g.][]{Starkenburg2014}. In addition, the comparison in
Fig.~\ref{fig:cumul_Pf} is based on the implicit assumption that our
models reproduce the observed CEMP fraction. However, the CEMP/VMP ratio
determined for the SAGA sample is larger than predicted by a factor of
$1.5$ up to $3$, depending on the model set (Table \ref{tab:SAGA}).
Hence, given all the uncertainties, the results of this comparison are
inconclusive.

%

\subsection{Masses of CEMP stars}
\label{mt}

\begin{figure}[!t]
\centering
\includegraphics[width=0.48\textwidth]{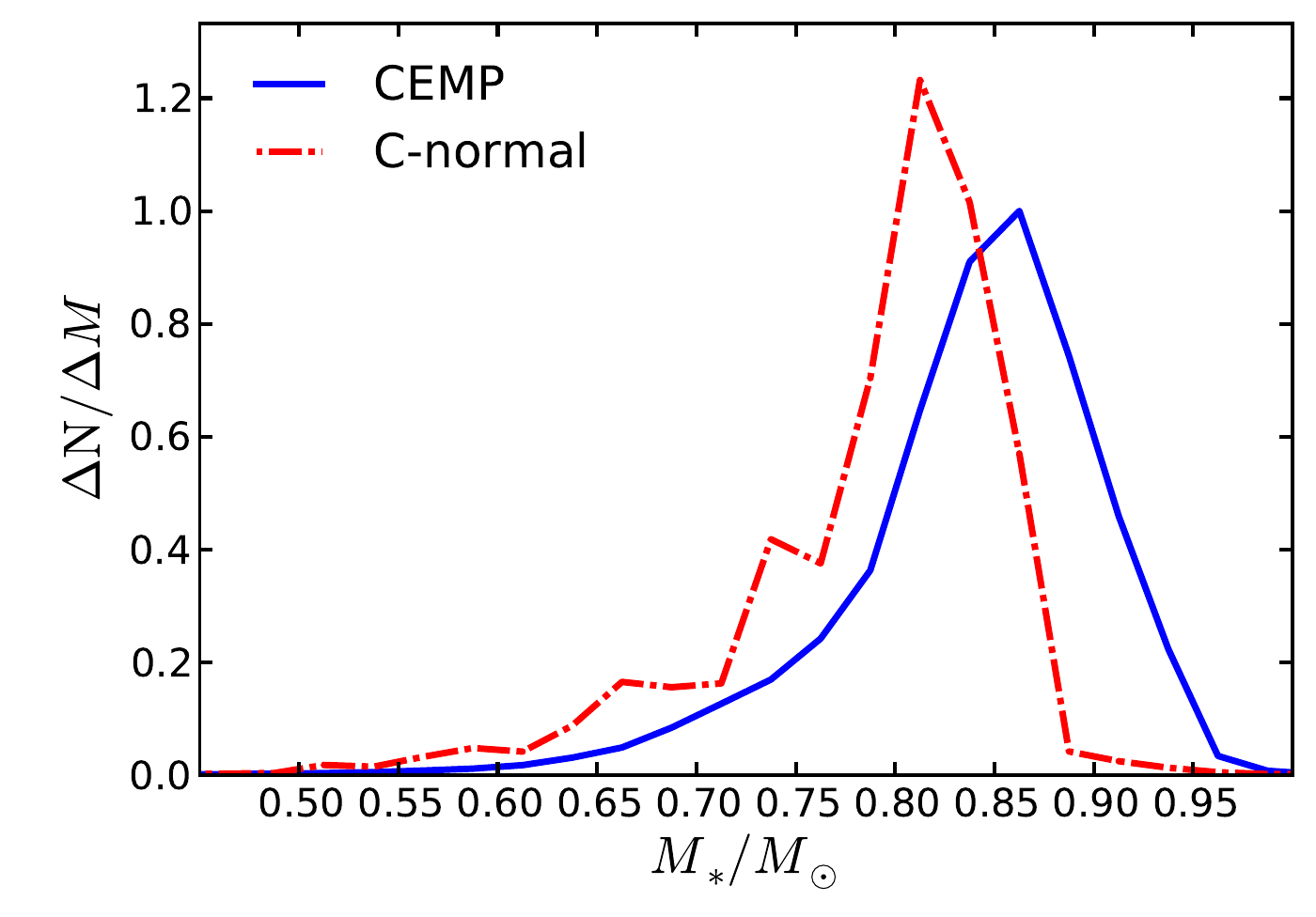}
\caption{Mass distribution of CEMP stars (solid line) and C-normal stars (dot-dashed line) computed with model set A. Bins of width $0.025$ dex are equally spaced in $M_*/\Msun$.}
\label{fig:mt}
\end{figure}

Our models predict that approximately $90\%$ of the CEMP stars have
masses between $0.75\,\Msun$ and $0.95\,\Msun$, with the peak of the
distribution around $0.85\,\Msun$. In Fig. \ref{fig:mt} we compare the
mass distributions computed with model set A for CEMP and carbon-normal
stars (solid and dot-dashed lines, respectively). The distribution of
CEMP stars is shifted towards higher mass by about $0.05\,\Msun$
compared to the distribution of carbon-normal stars. This difference is
a consequence of the mass-accretion process. A single star of mass
$0.85\,\Msun$ ascends the giant branch after approximately $10.8$
Gyr, and about $0.4$ Gyr later it becomes a white dwarf. By contrast,
if a star with an initial mass of $0.75\,\Msun$ accretes $0.1\,\Msun$
from a binary companion and becomes a CEMP star, its life is longer
than a normal $0.85\,\Msun$ star because until mass accretion
occurs the CEMP star evolves as a lower-mass star, i.e. more slowly.
Currently, no mass measurements of CEMP stars are available for
comparison with our models, but future observations that aim to detect very
metal-poor stars in eclipsing binary systems may allow the determination of
whether the predicted difference in mass is correct.

\begin{table*}[!t]
\footnotesize
\caption{Fractions of CEMP stars derived from the SAGA database and computed with our models.
The adopted $V$-magnitudes range in the models is $[6,\,16.5]$. CEMP stars have carbon abundance $[\C/\Fe]>+0.7$.
}
\label{tab:SAGA}
\centering
\begin{tabular}{c   r   r r r   r r r  }
\hline
\hline
	model sets and data	&	$\fC\,(\%)~~~$	&	$\fCgia\,(\%)~~$	&	$\fCTO\,(\%)~~$	&	$\fCdwa\,(\%)~~$	&	$\fgiaC\,(\%)$	&	$\fTOC\,(\%)$	&	$\fdwaC\,(\%)$	\\
\hline\\[-1.5ex]
SAGA database	&	$27.98\pm2.56$	&	$22.56\pm2.66$	&	$38.95\pm7.55$	&	$41.67\pm12.80$	&	$62.86\pm8.55$	&	$26.43\pm4.89$	&	$10.71\pm2.91$	\\[0.5ex]
\hline\\[-1.5ex]

A		&	$9.47 \pm 0.01$	&	$10.22 \pm 0.02$	&	$10.79 \pm 0.04$	&	$7.49 \pm 0.02$		&	$64.25 \pm 0.12$&	$12.17 \pm 0.04$	&	$23.58 \pm 0.06$	\\
B		&	$9.95 \pm 0.01$	&	$10.85 \pm 0.02$	&	$10.90 \pm 0.04$	&	$7.83 \pm 0.02$	&	$64.70 \pm 0.12$	&	$11.70 \pm 0.04$	&	$23.60 \pm 0.06$	\\
D		&	$11.67 \pm 0.04$	&	$10.57 \pm 0.05$	&	$15.04 \pm 0.14$	&	$12.64 \pm 0.08$	&	$53.94 \pm 0.30$	&	$13.77 \pm 0.13$	&	$32.29 \pm 0.22$	\\
$\IL$		&	$17.24 \pm 0.05$	&	$17.80 \pm 0.07$	&	$18.26 \pm 0.16$	&	$15.49 \pm 0.09$	&	$64.04 \pm 0.29$	&	$11.91 \pm 0.10$	&	$24.05 \pm 0.15$	\\
\hline
\end{tabular}
\end{table*}
	
%

\subsection{Abundance distributions}

The SAGA database includes the observed abundances of many elements. We
compare the abundance distributions derived from the observations with
those of our models. This comparison is qualitative because
the observed sample is inhomogeneous and incomplete, but it can provide
indications to constrain our models. The magnitude limits adopted in the
models are $\Vmin=6$ and $\Vmax=16.5$, consistent with the magnitude
range of the very metal-poor stars in the observed sample. In the first
part of this section we study the consequences of adopting different
masses of the partial mixing zone (Sect. \ref{PMZ}) while in the second
part we compare the predictions of different model sets (Sect.
\ref{abund}).

\subsubsection{The effect of $\Mpmz$}
\label{PMZ}
\begin{figure*}[!t]
\centering
\includegraphics[width=0.48\textwidth]{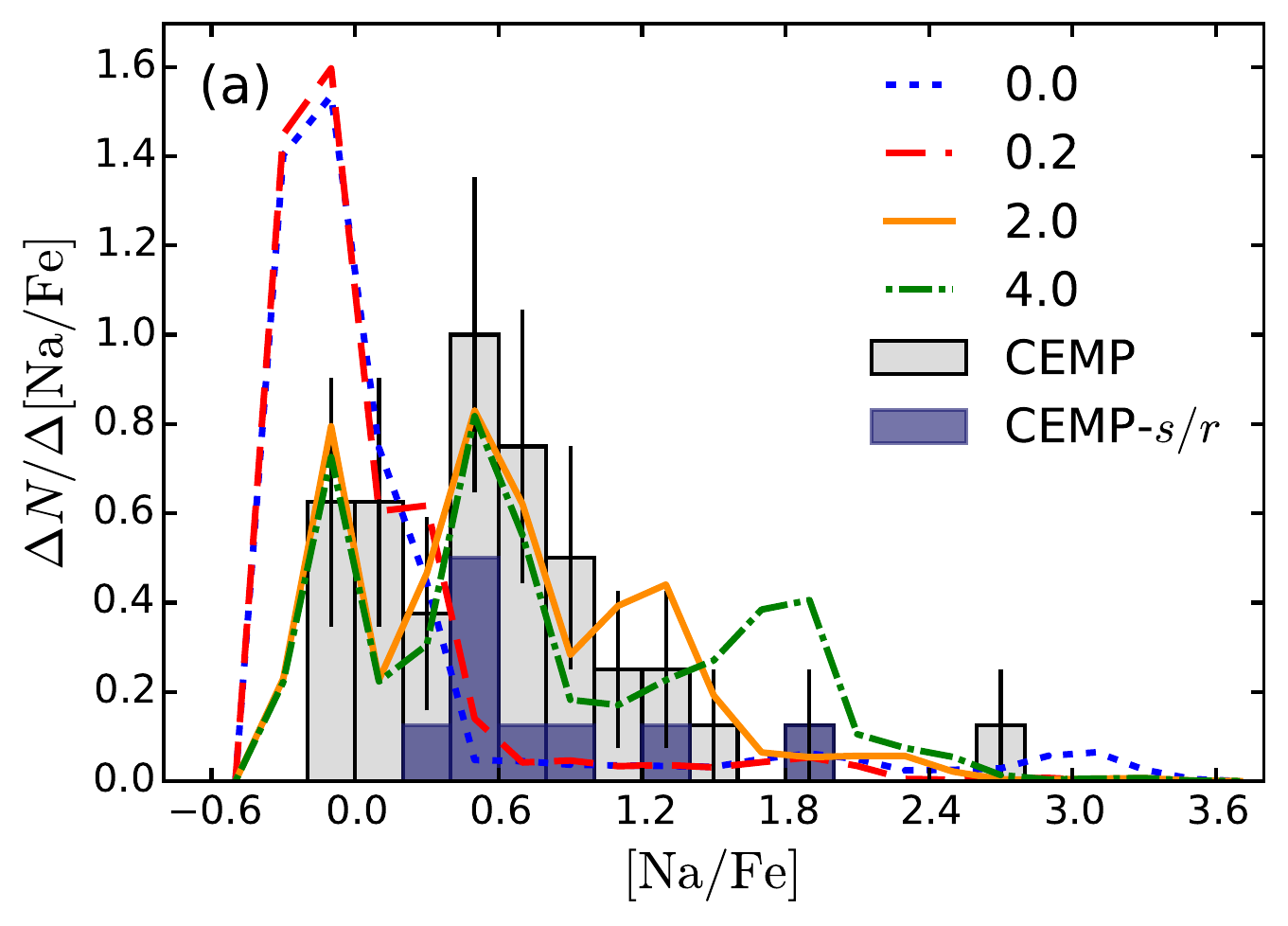}
\includegraphics[width=0.48\textwidth]{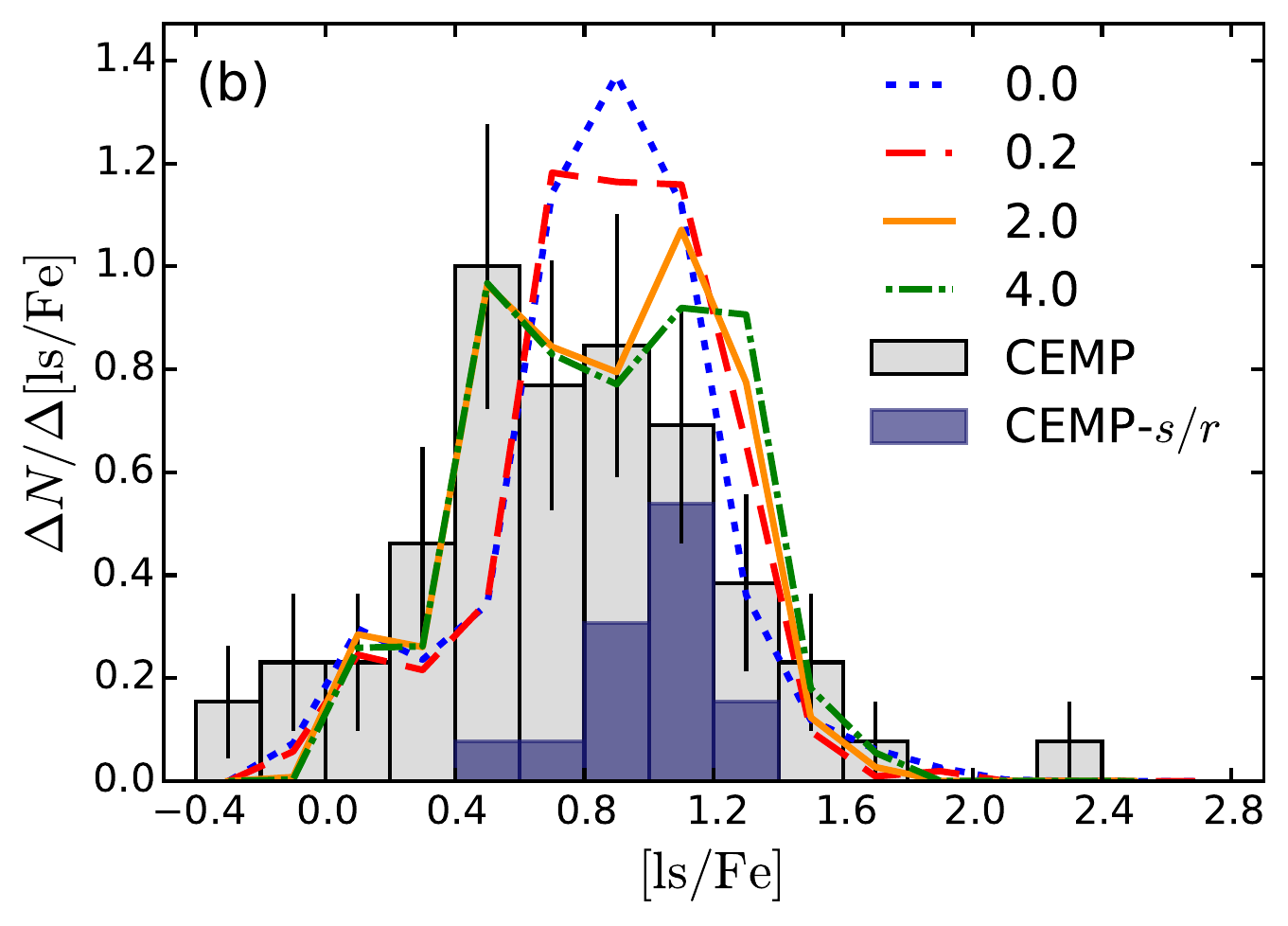}\\
\includegraphics[width=0.48\textwidth]{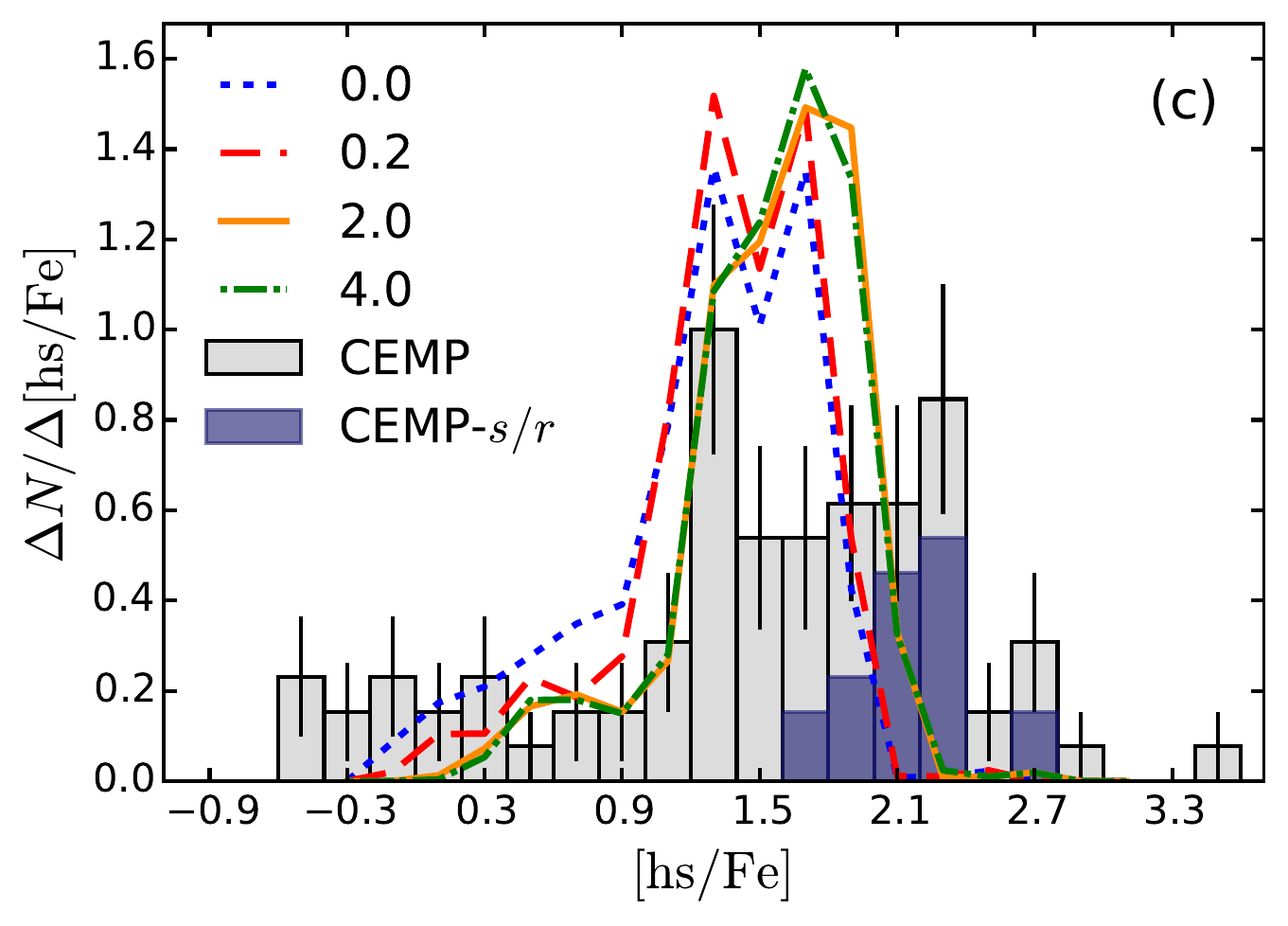}
\includegraphics[width=0.48\textwidth]{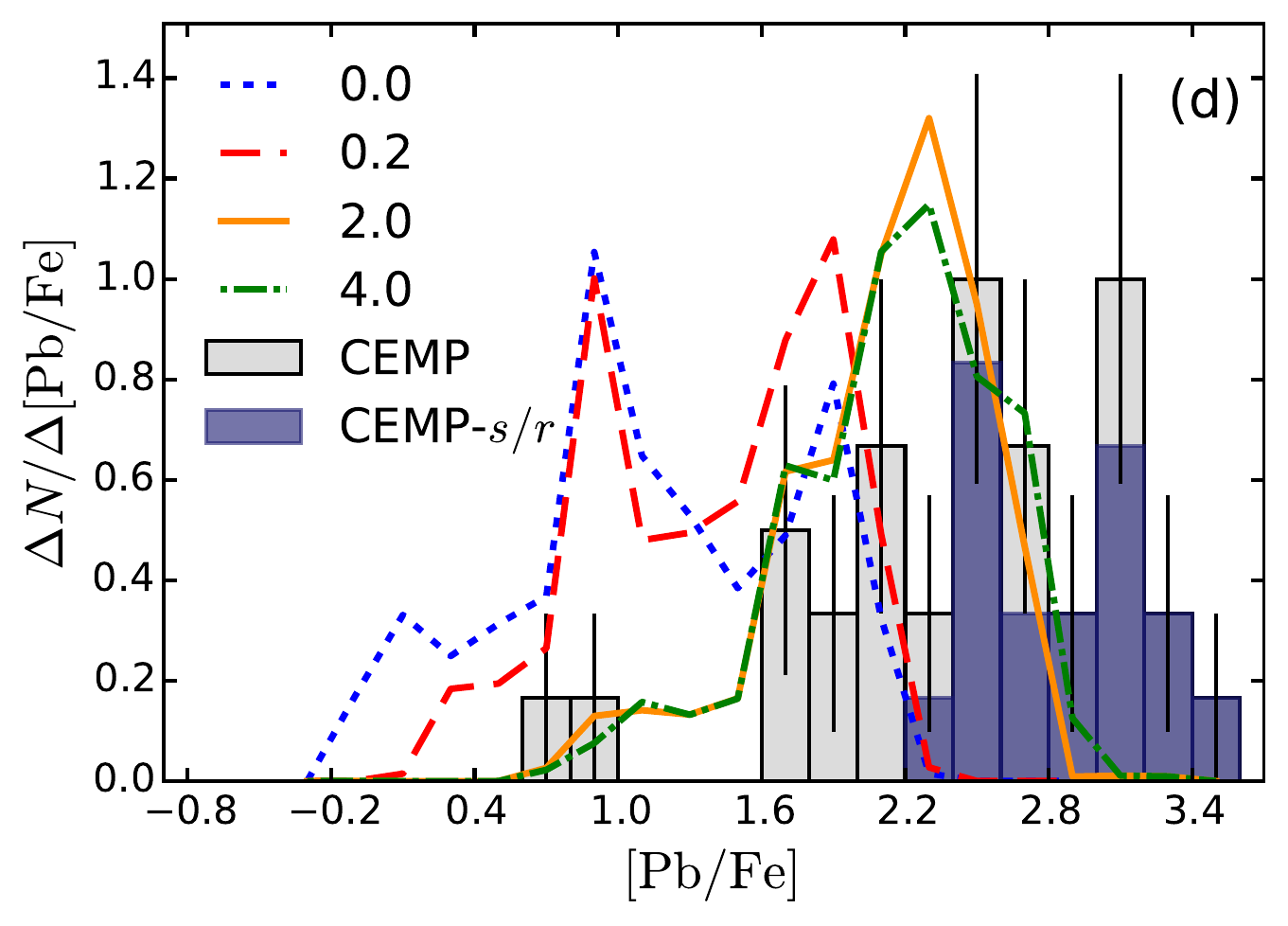}
\caption{Panels (a)--(d): abundance distributions of sodium, light-$s$ elements, heavy-$s$ elements and lead, respectively. The $0.2$ dex bins are equally spaced in [El/Fe]. The histograms represent CEMP and CEMP-$s/r$ stars from our database (light and dark grey, respectively), with Poisson errors. Models with $\Mpmz=0, 2\times10^{-4}\,\Msun$, $2\times10^{-3}\,\Msun$, $4\times10^{-3}\,\Msun$ are shown as dotted, dashed, solid, dot-dashed lines, respectively.
}
\label{fig:abunds_pmz}
\end{figure*}

The mass of the partial mixing zone is a free parameter in our models.
To understand its impact on the chemical properties of a population of
CEMP stars we compare the element distributions computed with default
model set A and four different values of $\Mpmz$: no PMZ, and $\Mpmz$
equal to $2\times10^{-4}\,\Msun$, $2\times10^{-3}\,\Msun$ and
$4\times10^{-3}\,\Msun$. Figs.
\ref{fig:abunds_pmz}a--d show the abundance
distributions of sodium, light-$s$ elements, heavy-$s$
elements,\footnote{ The abundances of light-$s$ and heavy-$s$ elements
are defined as [ls/Fe] = ([Sr/Fe]+[Y/Fe]+[Zr/Fe])/3 and [hs/Fe] =
([Ba/Fe]+[La/Fe]+[Ce/Fe])/3, respectively. If one of the elements is
missing, we average the abundances of the two available elements.
} %
and lead. The histograms represent the distributions of the observed
CEMP stars and the subsample of CEMP-$s/r$ stars (light and dark grey,
respectively). CEMP-$s/r$ stars are plotted separately because their
history of nucleosynthesis is probably different from CEMP stars with
low abundances of europium and $r$-elements \cite[e.g.][]{Jonsell2006,
Sneden2008, Lugaro2009, Bisterzo2012, Abate2015-1, Abate2015-2}. Our
models do not include the $r$-process, therefore in CEMP-$s/r$ stars we
do not expect to reproduce the abundances of elements with a strong
$r$-process component. However, the $r$-process is responsible for the
production of many neutron-capture elements, including some amounts of
elements traditionally associated with the $s$-process. In Paper II we
show that CEMP-$s/r$ stars have different abundance distributions than
$r$-normal CEMP-$s$ stars (i.e. with $[\Eu/\Fe]<+1$). In particular,
the abundances of neutron-rich elements such as barium, lanthanum,
cerium, and lead are typically strongly enhanced and the [hs/ls] ratio is
larger than is observed in $r$-normal CEMP-$s$ stars and than predicted
by our models, in line with the results of previous studies \cite[e.g.
][]{Bisterzo2012, Lugaro2012}.

\begin{figure*}[!t]
\centering
\includegraphics[width=0.48\textwidth]{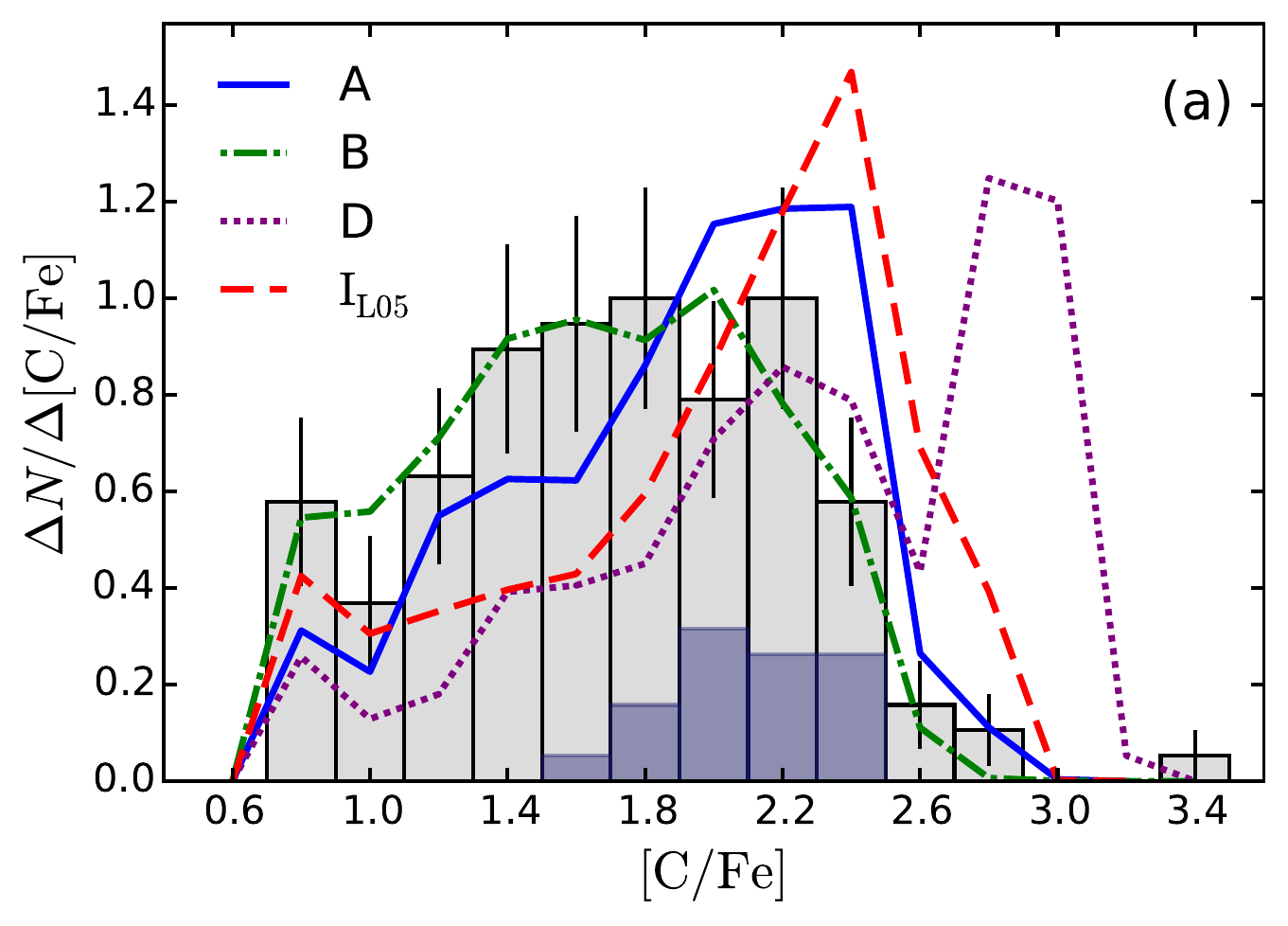}
\includegraphics[width=0.48\textwidth]{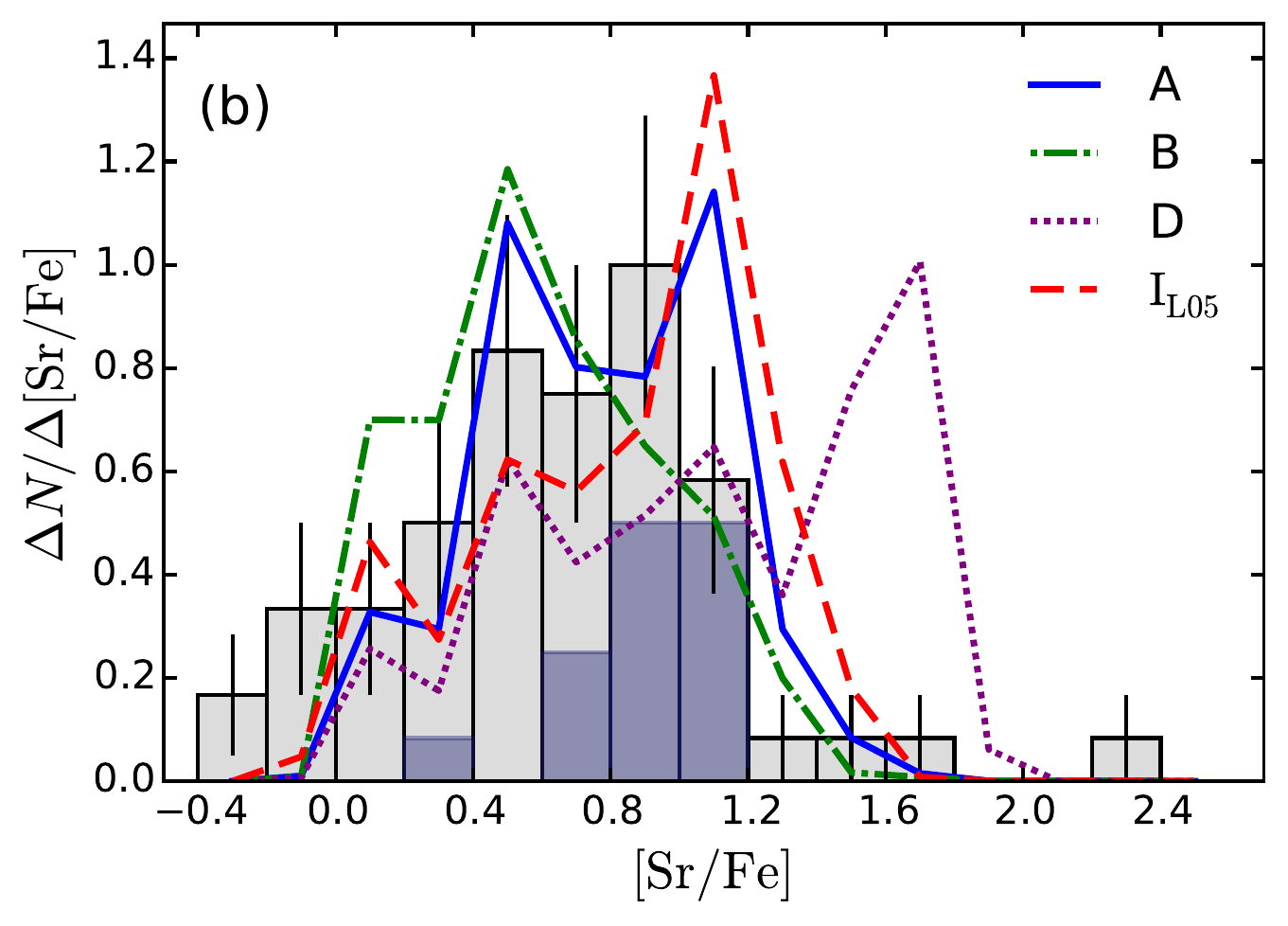}\\
\includegraphics[width=0.48\textwidth]{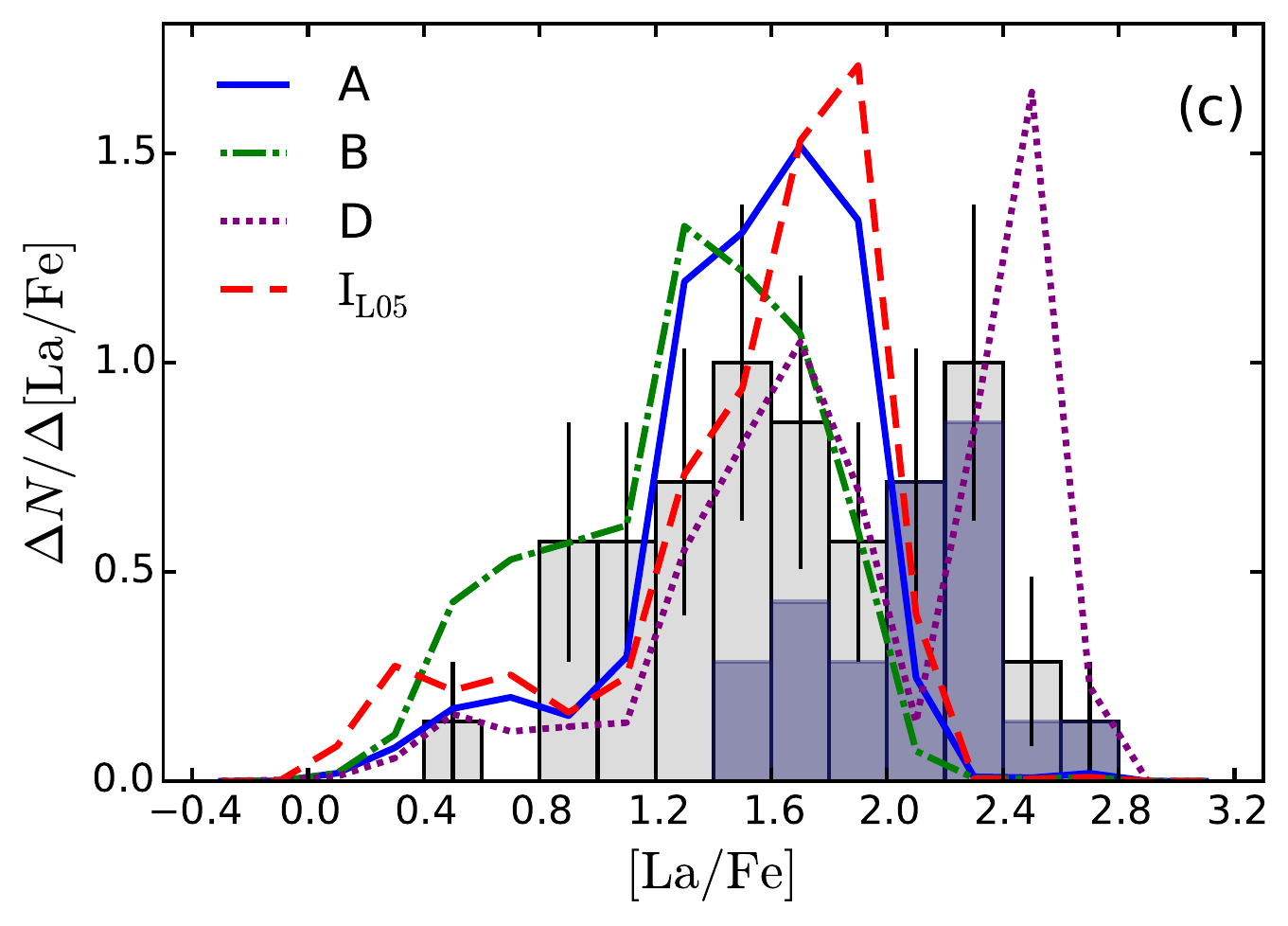}
\includegraphics[width=0.48\textwidth]{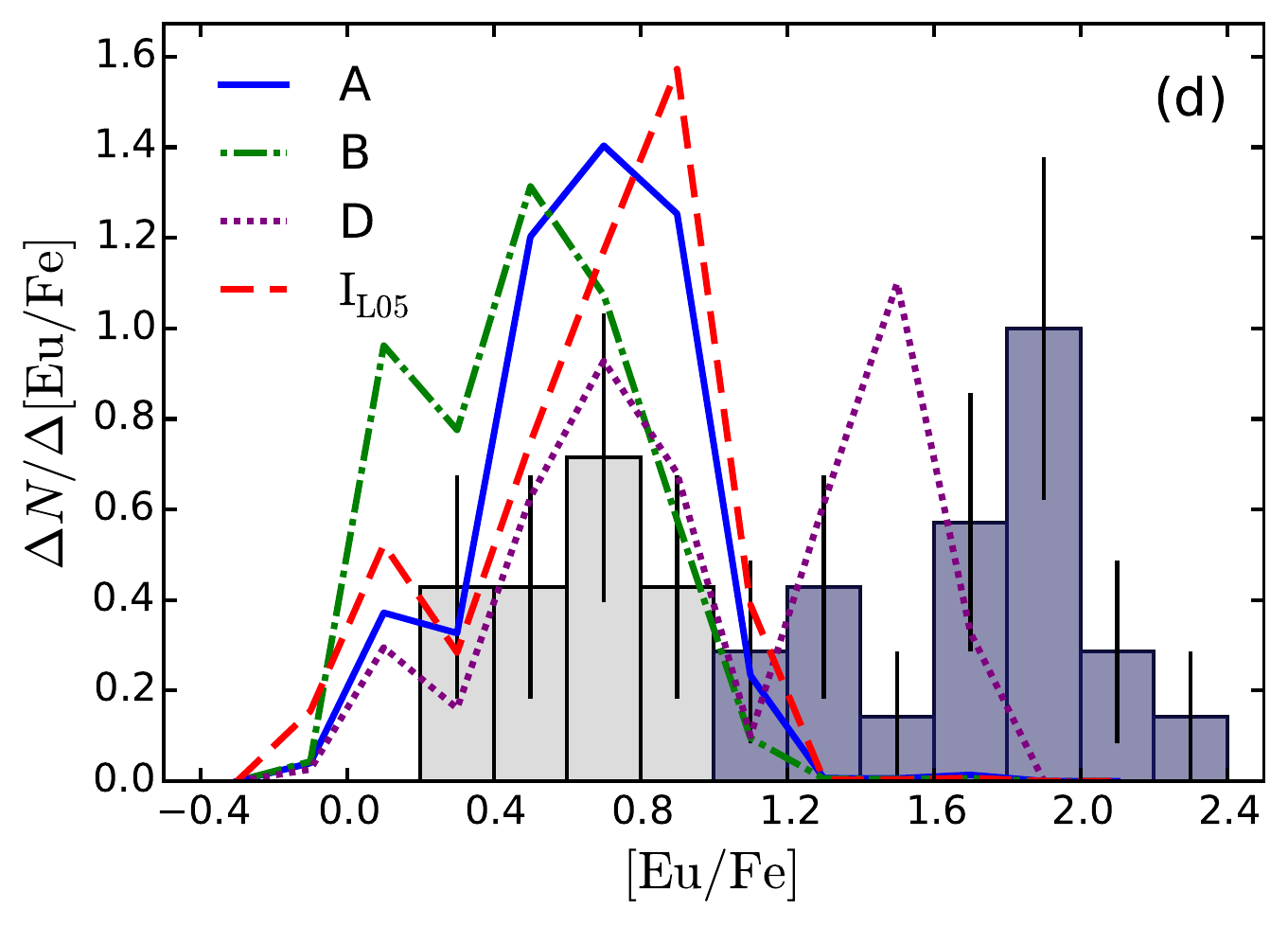}
\caption{Panels (a)--(d): distributions of $[\C/\Fe]$, [Sr/Fe], [La/Fe],
and [Eu/Fe], respectively. Model sets A, B, D, and $\IL$ are shown with
solid, dot-dashed, dotted, and dashed lines, respectively. Light and dark grey
histograms represent the observed CEMP and CEMP-$s/r$ stars in the SAGA database.
The $0.2$ dex bins are equally spaced in [El/Fe].}
\label{fig:abunds}
\end{figure*}

In the models with a small PMZ ($\Mpmz=0$ and $2\times10^{-4}\,\Msun$),
the majority of CEMP stars have normal, non-enhanced abundances of sodium
(Fig. \ref{fig:abunds_pmz}a). This is because the abundance of sodium
produced in AGB nucleosynthesis is sensitive to both the mass of the
star and of its PMZ. With no PMZ, or only a small PMZ, only relatively massive AGB
stars produce significant amounts of sodium. In the modelled population,
most CEMP stars are formed in systems with a low-mass primary star,
hence the distribution peaks around the initial abundance. On the other
hand, with a relatively large PMZ ($\Mpmz=2-4\times10^{-3}\,\Msun$),
low-mass AGB stars also produce significant amounts of sodium;
consequently the modelled distribution qualitatively reproduces the
observations. 

The distributions of light-$s$ and heavy-$s$ elements (Figs.
\ref{fig:abunds_pmz}b and \ref{fig:abunds_pmz}c) do not vary
significantly as a function of $\Mpmz$, because low-mass AGB stars
(below $\approx2\,\Msun$) experience episodes of proton ingestion from
the convective envelope in the He-flash induced convection zone
\cite[][]{Lugaro2012}, and some amounts of $s$-elements are produced
regardless of the PMZ\footnote{%
The nucleosynthetic signature of a proton-ingestion event is
uncertain because one-dimensional stellar-evolution models do not adequately
treat the physics of these inherently three-dimensional phenomena
\cite[see e.g.][]{Stancliffe2011, Herwig2014}.
}. %
Because low-mass primary stars dominate our
synthetic binary population, the distributions of light-$s$ and
heavy-$s$ elements in CEMP stars are quite similar. By contrast, the
abundance of lead is related more strongly to the PMZ mass, as shown in
Fig. \ref{fig:abunds_pmz}d, and the distributions obtained with small
and large $\Mpmz$ are shifted by about $0.5$ dex. Models with
$\Mpmz\ge2\times10^{-3}\,\Msun$ are in better agreement with the
observed distribution; however, none of the models can reproduce the
$11$ observed CEMP stars with $[\Pb/\Fe]\ge +2.8$. Nine of these stars are
CEMP-$s/r$ stars, and the discrepancy with the models may indicate that
lead in these extremely enriched stars was not produced purely by the
$s$-process. %
We note that the observed abundances are determined assuming
local thermodynamic equilibrium. It has been shown that departures from
this approximation can greatly impact the abundances of giants and 
metal-poor stars \cite[e.g.][]{Bergemann2012-3}. In particular, positive corrections
of 0.3--0.5~dex are to be expected for the lead abundances of stars
in the metallicity range of our observed sample \cite[][]{Mashonkina2012}. If we assume an
average correction of $0.4\,$dex, $21$ CEMP-$s$ stars (of which $16$ are CEMP-$s/r$)
would have $[\Pb/\Fe]\ge2.8$ and cannot be reproduced by any of our models.

The results of the comparison between models with different partial mixing zones,
as well as the results of Papers I and
II, suggest that to reproduce the chemical properties of the observed
sample our models require a relatively large $\Mpmz$. Therefore, in the
following we adopt $\Mpmz=2\times10^{-3}\,\Msun$ in our model sets.

\subsubsection{Abundance distributions with different model sets}
\label{abund}

Our modelled abundance distributions depend mostly on the distributions
of the primary and secondary stellar mass of our synthetic CEMP stars;
the former determines the amount of each element produced by AGB
nucleosynthesis, while the latter essentially determines how much mass
is accreted and consequently the abundance enhancement. We compare the
results of our default model set A with three model sets: model set B,
in which the $\Msec$ distribution of CEMP stars is more peaked towards
relatively massive secondary stars ($\Msec>0.7\,\Msun$), model set
$\IL$, in which a significant proportion of CEMP stars are formed in
binary systems with a primary star initially above $1.2\,\Msun$, and
model set D, in which thermohaline mixing is inhibited and therefore the
abundances of the accreted material are not diluted in turnoff and dwarf
CEMP stars. The CEMP fractions observed in the SAGA database and
computed with these four model sets are shown in Table \ref{tab:SAGA}.
In model sets A, B, and $\IL$, the CEMP/VMP ratios are higher
than the corresponding sets in Table \ref{tab:SEGUE} because for the
comparison with SAGA stars we adopt lower magnitude limits in our
simulations, thus the fainter dwarf stars are selected against. These
fainter stars have a larger proportion of carbon-normal stars, so the
CEMP/VMP ratios increase. On the other hand, in model set D, among the
fainter dwarf stars the CEMP stars are more frequent because of the lack
of dilution, thus $\fC$ decreases.

Figures \ref{fig:abunds}a--\ref{fig:abunds}d show the abundance
distributions of carbon, strontium, lanthanum, and europium. We find
similar trends for all elements. Compared to the default model A, model
set $\IL$ favours more massive primary stars, which typically produce
larger abundances of carbon and $s$-elements for $\Mpmz=2\times10^{-3}\,
\Msun$. Also, in model set $\IL$ more CEMP stars are formed with
$\Msec\le0.6\,\Msun$, which accrete large amounts of material from the
primary star. Consequently, the abundance distributions of model set
$\IL$ are weighted towards large enhancements. On the contrary, model
set B favours relatively massive secondary stars, $\Msec\ge0.7\,\Msun$,
which accrete small amounts of material that is more strongly diluted.
Consequently, model set B predicts abundance distributions that peak at
lower [El/Fe] than our default model A. With model set D, dwarfs form a
clearly distinct group of stars. Because the accreted material is not
efficiently mixed throughout the entire star, these stars have a
chemical composition very similar to that of the polluting stars, and
hence the largest enhancements.  

The distributions of lanthanum and europium (respectively, a heavy-$s$
and an $r$-element) in Figs. \ref{fig:abunds}c and \ref{fig:abunds}d
show that the abundances observed in CEMP-$s/r$ stars are outside the
range that can be produced by our models. If we exclude the CEMP-$s/r$
stars from the sample, our models qualitatively reproduce the observed
distributions. This discrepancy is observed also in the distributions of
[hs/Fe] and [Pb/Fe] (Figs. \ref{fig:abunds_pmz}c and
\ref{fig:abunds_pmz}d). By contrast, the distributions of carbon,
sodium, strontium, and light-$s$ elements do not differ significantly
between CEMP-$s/r$ stars and $r$-normal CEMP-$s$ stars. These results suggest
that there is a relation between the strongly enhanced abundances of the
heavy-$s$ elements and the $r$-elements in the population of CEMP-$s/r$
stars, in line with our conclusions in Paper~II.

\begin{figure}[!t]
\centering
\includegraphics[width=0.48\textwidth]{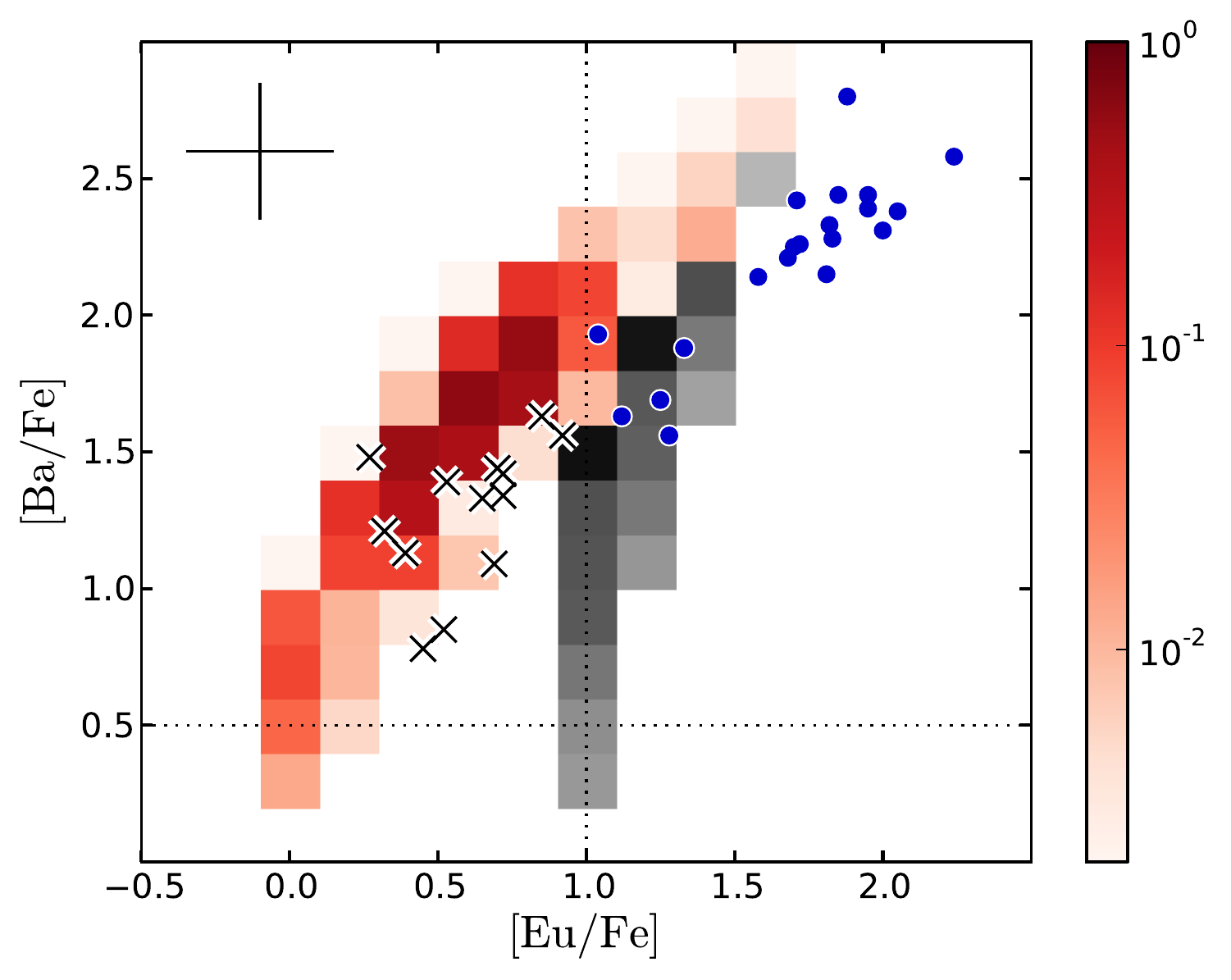}
\caption{Distribution of $[\Ba/\Fe]$ vs. $[\Eu/\Fe]$. The red distribution represents our default model A. The grey distribution is computed with an initial enhancement of $[r/\Fe]_{\mathrm{ini}}=1$. The dotted lines indicate the thresholds $[\Ba/\Fe]=+0.5$ and $[\Eu/\Fe]=1$ that define CEMP-$s$ and CEMP-$s/r$ stars, respectively. The crosses indicate the observed CEMP-$s$ stars with $[\Eu/\Fe]<+1$. CEMP-$s/r$ are shown as blue circles. The +~symbol in the top-left corner shows the average observed uncertainty.}
\label{fig:BaFeEuFe}
\end{figure}

The origin of the $r$-enhancement in CEMP-$s/r$ stars is uncertain. If
we start our simulations with an initially enhanced abundance of
$r$-elements to mimic the effect of a primordial enrichment independent
of the abundances of $s$-elements, as suggested e.g. by
\cite{Bisterzo2012}, our models fail to reproduce the stars with the
lowest [Eu/Fe] without improving the results obtained for CEMP-$s/r$
stars. Fig. \ref{fig:BaFeEuFe} shows the observed abundances of europium
and barium in the CEMP stars of our sample, together with the synthetic
abundances computed with our default model set A (red distribution),
assuming an initial enhancement of $[r/\Fe]_{\mathrm{ini}}=1$ (grey
distribution). The observations exhibit a clear linear correlation
between the abundances of barium and europium. Our model set A
qualitatively reproduces this correlation for $r$-normal CEMP-$s$ stars,
that is, to the left of the vertical dotted line at $[\Eu/\Fe]= +1$. As
discussed by \cite{Lugaro2012}, an initial enhancement of
$[r/\Fe]_{\mathrm{ini}}\le0.4$ dex is essentially washed out by the
europium produced by the $s$-process in the AGB nucleosynthesis. If we
adopt $[r/\Fe]_{\mathrm{ini}}= +1$, europium increases while barium does
not change, and the observed correlation is not reproduced. If we adopt
$[r/\Fe]_{\mathrm{ini}}= +2$ to reproduce the most $r$-rich CEMP-$s/r$
stars, then the abundances of the $r$-elements in our synthetic CEMP
stars all end up in one bin around this value.

\begin{figure}[!t]
\centering
\includegraphics[width=0.48\textwidth]{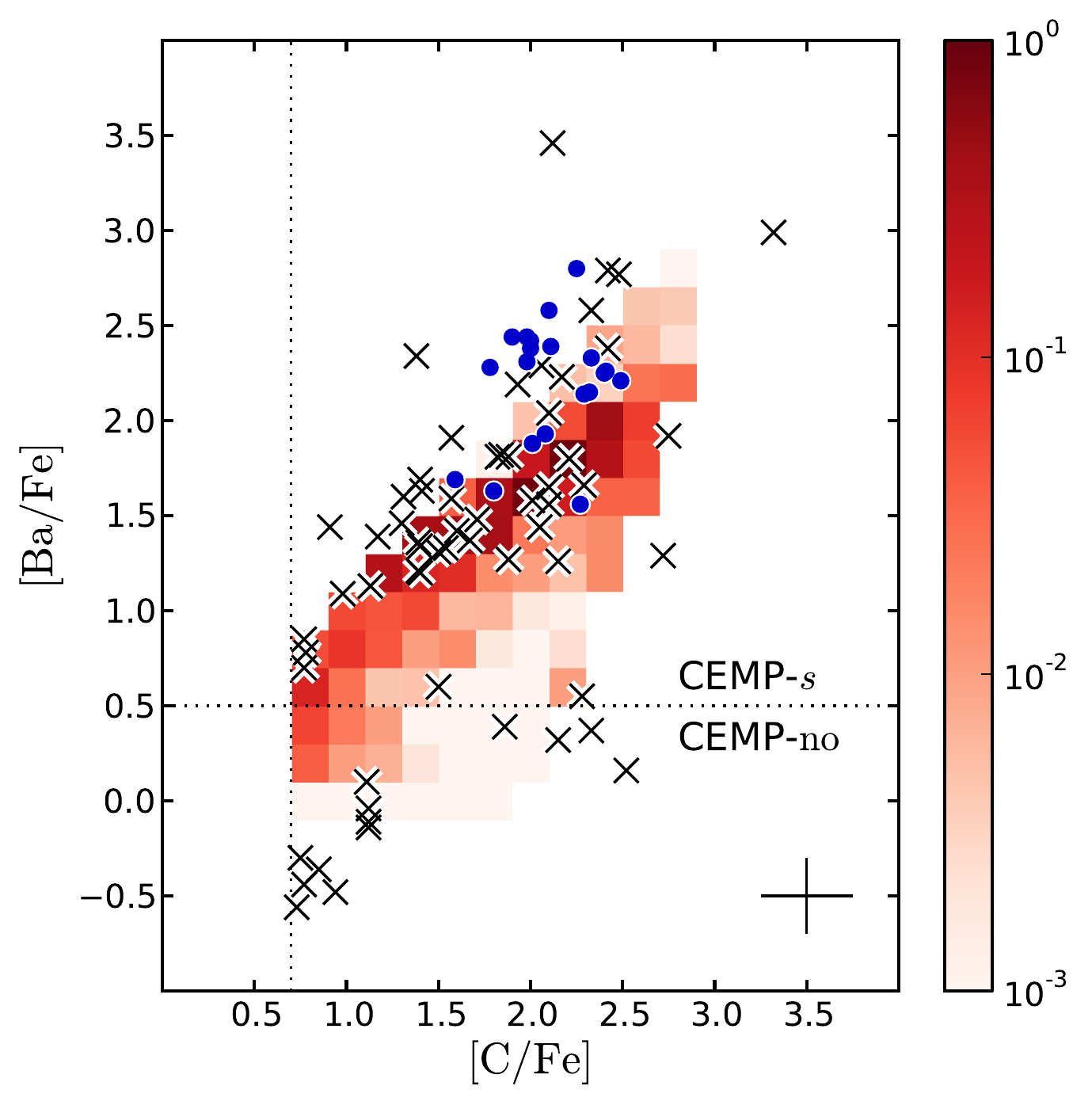}
\caption{As Fig. \ref{fig:BaFeEuFe} for $[\Ba/\Fe]$ vs. $[\C/\Fe]$. The vertical line represents the threshold $[\C/\Fe]=+0.7$. In the bottom-right corner the average observed uncertainty is shown as a +~symbol.}
\label{fig:CFeBaFe}
\end{figure}

Figure \ref{fig:CFeBaFe} shows the distribution of [Ba/Fe] vs. [C/Fe]
computed with model set A. An apparent correlation exists between carbon
and barium in the observed CEMP-$s$ stars, regardless of the
$r$-enhancement, although there is considerable scatter in the data. In
the binary scenario for the formation of CEMP stars this correlation is
a direct consequence of nucleosynthesis in AGB stars. The amounts of
carbon and barium produced by an AGB star are correlated, and depend
only on the stellar mass (for a fixed PMZ). The observed correlation is
qualitatively reproduced by our model, although a fraction of the CEMP
stars have a barium abundance systematically larger than the model
predictions. This discrepancy is not removed even with higher masses of
the PMZ, and suggests that, for a given amount of carbon, our AGB models
should produce a larger amount of barium.

The observed correlation between carbon and barium breaks down for CEMP-no stars,
whose carbon abundances are spread over two orders of magnitude, and are
not well reproduced by our models. This discrepancy may arise within the
mass-transfer scenario because a large abundance of carbon,
combined with a small abundance of barium, can be obtained in models of
massive AGB stars ($M>2.8\Msun$) with no PMZ, and massive stars are
relatively rare in our population. However, about half of the CEMP-no
stars in our sample have $[\Ba/\Fe]<0$, and with our model it is not
possible to reproduce their abundances. The formation scenario of
CEMP-no stars is currently uncertain. \cite{Masseron2010} show that
there is a clear continuity from CEMP-$s$ to CEMP-no in the abundance
trends of several elements, including C+N, O, Mg, $^{12}\C/\Cth$ and
C/N, pointing towards the scenario of mass transfer from an AGB
star. On the other hand, the study of \cite{Starkenburg2014}
indicates that CEMP-no stars are not found in binary systems more often
than carbon-normal stars or, alternatively, they belong to very wide
binaries. Although the hypothesis of wide separations is consistent with
our results, the evidence that with our models most of the CEMP-no stars
are not reproduced in the carbon--barium space may be an indication that
their observed carbon enhancement is not related with AGB
nucleosynthesis but rather is the result of contributions from
first-generation stars.


\section{Discussion}
\label{discussion}

In this paper we have simulated populations of binary stars at metallicity
$Z=10^{-4}$ ($[\Fe/\Hy\approx-2.3$) to investigate
the mass-transfer scenario for the formation of CEMP stars. A comparison
with similar studies \cite[e.g.][and \citealp{Abate2013}]{Izzard2009}
shows that our default model set A predicts a CEMP/VMP ratio more than
three times larger than previous models adopting default initial assumptions.
In the default model adopted in \cite{Abate2013}, AGB stars of mass
below $1.2\,\Msun$ do not undergo third dredge-up, whereas in our models
\cite[based on the detailed calculations of][]{Karakas2010} the minimum
mass for third dredge-up in AGB stars is $0.9\,\Msun$. Therefore, the
range of primary masses that contribute to the formation of CEMP stars
is larger and, as a consequence, the CEMP/VMP ratio predicted in our
models increases. 
In our simulations we adopt a selection criterion based on the
luminosity of the stars, instead of a simple cut-off in surface gravity
as in \cite{Abate2013}. As a result, the proportion of dwarf stars in
the CEMP population increases to approximately $50\%$, but the total
CEMP/VMP ratio remains essentially the same. 

If we adopt a solar-neighbourhood IMF, our CEMP/VMP ratio, $\fC$, is
approximately $8\%$ for $[\C/\Fe]>+1$, consistent with the observed
value of the SEGUE sample in the range $-2.5<[\Fe/\Hy]\le-2.0$.
Therefore, a change of the IMF towards relatively large primary masses
(as suggested, e.g. by \citealp{Lucatello2005a}, and \citealp{Suda2013})
is not necessary to reproduce the observed CEMP/VMP ratio. However, the
proportion of dwarf CEMP stars predicted in our models is much larger
than observed in the SEGUE sample at $-2.5<[\Fe/\Hy]\le-2.0$. Also, the
total number of dwarf stars observed at this metallicity in the SEGUE
sample is a factor of four to five lower than the number of turnoff and
giant stars. These differences may suggest that our simulations include
stars that in reality are too faint to be detected, or that the observed
surface gravities are possibly underestimated by a few tenths of a dex. 

The CEMP/VMP ratio observed in the SEGUE sample increases by almost a
factor of two if the minimum carbon abundance in the definition of CEMP
stars is reduced from $[\C/\Fe]=+1.0$ to $[\C/\Fe]=+0.7$. By contrast,
this causes only a small difference in our models.
In Paper~II we showed that most carbon-normal VMP stars from the SAGA database exhibit carbon
abundances within approximately $0.4$ dex from the average value of
$[\C/\Fe]=+0.3$. Assuming that carbon-normal VMP stars in the SEGUE
sample have the same distribution of carbon abundances, and considering
that an observational error of $0.3$ dex is associated with carbon
abundances in SEGUE stars, one explanation of the large increase in the
observed CEMP fraction between $[\C/\Fe]=+0.7$ and $[\C/\Fe]=+1.0$ is
that we are observing the tail of the carbon-abundance distribution of
carbon-normal very metal-poor stars. 

The CEMP/VMP ratio calculated among stars in the SAGA database is
very different from the SEGUE sample. In the range
$-2.8\le[\Fe/\Hy]\le-1.8$ we find $\fC=28\%$, adopting
$[\C/\Fe]_{\mathrm{min}}=+0.7$, almost a factor of three larger than the
value calculated for SEGUE stars. A discrepancy of about a factor of
three is also found with most of our model sets, which predict CEMP/VMP
ratios around $10\%$. Another remarkable difference with the SEGUE
sample is that the CEMP fraction from the SAGA sample increases towards
stars of higher $\logg$. A possible explanation of these discrepancies
is that the SAGA sample is a compilation of all very metal-poor stars
currently available in the literature. Hence, it is an inhomogeneous and
incomplete sample, and is possibly biased towards chemically peculiar
stars. However, the large differences between the results obtained with
the SEGUE and SAGA samples point out the general problem that the
measured fractions depend significantly on the observed sample
considered. Consequently, the results of the comparisons with population
synthesis models are as yet inconclusive.

In our simulations we assume a binary fraction of unity in the range of
orbital separations between $5\,\Rsun$ and $5\times10^6\,\Rsun$. For
comparison, \cite{Kouwenhoven2007} find that, with $3\sigma$ confidence,
the binary fraction in the young stellar association Scorpius OB2 is at
least $0.7$, and probably close to unity for stars of spectral type A
and B in the same range of separations. However, the binary fraction of
the stellar population of the Galactic halo is poorly constrained
\cite[][]{Carney2001, Carney2005, Rastegaev2010, Gao2014, Aoki2015}.
Recent results suggest that among stars with $[\Fe/\Hy]<-1.1$ the binary
fraction is at least $50\%$ and possibly increasing with decreasing
metallicity \cite[][]{Gao2014, Yuan2015-2}.

All our model sets predict that CEMP stars are formed in initially wide
binary systems with orbital periods typically longer than $1,\!000$ days
up to approximately $10^6$ days. About $80\%$ of the simulated CEMP
stars have periods longer than $5,\!000$ days, which is approximately
the longest period currently observed. The median of the observed period
distribution is approximately $500$ days, consistent with the average
orbital period estimated by \cite{Starkenburg2014} for a different
sample of CEMP-$s$ stars, whereas for the synthetic distributions it is
about $20,\!000$ days. This indicates that, if the distribution
determined from the CEMP stars with measured orbital periods is
representative of the entire CEMP population, our simulations
significantly underestimate the observed CEMP/VMP ratio. A similar
discrepancy between the observed periods and model predictions also
exists at higher metallicity for the barium stars, which are likely
formed by the same wind-accretion process as CEMP stars
\cite[e.g.][]{BoffinJorissen1988, Jorissen1998}. For barium stars, the
observed sample of orbital periods is complete, and only a few systems
are known with periods longer than $10^4$ days. The discrepancy between
the observed and modelled periods suggests that in our models binary
systems need to efficiently lose angular momentum, and transfer mass
more efficiently at short separations, as also discussed by \cite{Izzard2010}
and in Papers~I and II.

The comparison between the observed and synthetic abundance
distributions provides constraints on our nucleosynthesis model and on
the properties of the progenitor systems. The chemical composition of a
modelled CEMP star depends essentially on the mass of the primary star,
which determines the elements produced by AGB nucleosynthesis, and on
the initial mass of the secondary star, which determines the amount of
material that needs to be accreted so that after $10$ Gyr the star is
luminous enough to be visible and enriched in carbon. Most model sets
predict similar distributions of $\Mprim$ and $\Msec$, consequently the
abundance distributions are similar. 

The modelled abundance distributions of most elements do not show a
strong dependence on the mass of the partial mixing zone, $\Mpmz$. The production
of $s$-elements is sensitive to $\Mpmz$, especially for stars with
$\Mprim\ge1.5\,\Msun$, while most CEMP stars in our simulations are
formed from binary systems with $\Mprim\le1.2\,\Msun$. 
By contrast, in Papers~I and II we find that primary masses above $1.4\,
\Msun$ are necessary to reproduce the detailed chemical composition of
most observed CEMP-$s$ stars of our sample. Systems with primary mass
between $1.5\,\Msun$ and $2.5\,\Msun$ are required to reproduce the
element-to-element ratios observed in CEMP-$s$ stars that exhibit
abundant heavy-$s$ elements and large [hs/ls] and [Pb/hs] values. These
chemical properties are observed mostly in CEMP-$s/r$ stars, and also in
some $r$-normal CEMP-$s$ stars. The discrepancy between the results of
the population-synthesis simulations and the detailed analysis of Paper
I and II may hint that nucleosynthesis models of low-mass AGB stars
($M\lesssim1.2\,\Msun$) should produce, in some circumstances, higher
abundances of heavy-$s$ elements and lead. 

We note that the observed abundances are normally determined
under the assumption of local thermodynamic equilibrium, which is not
always fulfilled in the layers of the stellar atmosphere where spectral 
lines are formed \cite[e.g.][]{Mihalas1973}. The departure from local thermodynamic
equilibrium in real stars introduces a bias in the abundance determination of
most elements, including carbon, iron, strontium, barium, and lead. For example,
\cite{Mashonkina2012} show that a positive correction of
about 0.3--0.5~dex for metal-poor stars should be applied
to the observed lead abundances. This correction increases for increasing temperature,
and for decreasing metallicity and surface gravity. Positive corrections are typically
required also for barium and europium \cite[][]{Mashonkina2012, Bergemann2014}. 
Although the magnitudes of the corrections are of the order of the observational
uncertainties ($\lesssim0.2$\,dex), they suggest that our models should produce
higher abundances of heavy elements for a given amount of carbon.

Our default model set A predicts abundance distributions of light
elements (e.g. carbon and sodium) and light-$s$ elements that are
qualitatively consistent with the observed distributions in all CEMP
stars. 
In contrast, the abundance distributions of heavy-$s$ elements, lead, and
europium are reproduced only in $r$-normal CEMP stars, whereas in
CEMP-$s/r$ stars the abundances of these elements are outside the range
of our model predictions. 
Similarly, a clear correlation between the abundances of barium and
europium is observed in all CEMP-$s$ stars. Our model reproduces such a
correlation in CEMP stars with $[\Eu/\Fe]<+1$, while in CEMP-$s/r$ stars
the observed enhancements of barium and europium are too large.
Consistent results are found in the detailed analysis of CEMP-$s$ stars
performed in Papers~I and II. In Paper~I we find that to reproduce the
large abundances of heavy-$s$ elements observed in CEMP-$s/r$ stars, our
models typically overestimate the abundances of carbon, sodium,
magnesium, and light-$s$ elements. Paper~II shows that in $r$-normal
CEMP-$s$ stars the abundances of most elements are reproduced on average
within the observed uncertainty, whereas in CEMP-$s/r$ stars the
abundances of both heavy-$s$ and $r$-elements are systematically
underestimated. 

Although enhanced abundances of $r$-elements are also observed in
carbon-normal stars with no evidence of duplicity (e.g. \citealp{Barklem2005};
\citealp{Roederer2014-3, Roederer2014-1}; Hansen et al. {\it in prep.}),
our results suggest that the $s$- and $r$-processes responsible for
the abundances observed in CEMP-$s/r$ stars may have occurred in the
same astrophysical site. Consequently, these results indicate that under
some conditions AGB stars may be able to reach the large densities
necessary to activate the $r$-process, in contrast to what is currently
found in the AGB models. Detailed models of AGB stars at extremely low
or zero metallicity show that large neutron densities
\cite[$10^{12}-10^{16}\cm^{-3}$,][]{Campbell2008, Herwig2011} may be
reached as a result of proton ingestion in the helium-flash region.
\cite{Lugaro2009} propose a speculative scenario in which extremely
low-metallicity ($Z<10^{-5}$) AGB stars may be able to produce both $s$-
and $r$-elements. Further evolutionary and nucleosynthetic calculations
are needed to test quantitatively if this scenario may also work at
metallicity $Z\approx10^{-4}$, and if the predicted abundances reproduce
the chemical compositions of observed CEMP-$s/r$ stars.

%


\section{Conclusions}
\label{concl}

In conclusion, in our synthetic population of very metal-poor binary
stars we find a CEMP fraction that varies between $7\%$ and $17\%$,
depending on the initial assumptions. Our default model set predicts a
CEMP/VMP fraction of $8.5\%$ for $[\C/\Fe]>+0.7$. This fraction is more
than three times larger than the results obtained by \cite{Izzard2009}
and \cite{Abate2013} with the same population-synthesis code and default initial assumptions. This
difference results from the updates that we included in the model of AGB
nucleosynthesis. With the default models adopted in the previous work, AGB stars below $1.2\,\Msun$ did
not undergo third dredge-up, whereas in our updated models AGB stars
experience third dredge-up down to $0.9\,\Msun$. Hence, the range of
primary masses that can form a CEMP star is increased.

The observed fraction of CEMP stars depends significantly on the sample
that is taken into account. In the most recent study of the SDSS/SEGUE
stellar survey a CEMP fraction of $11.5\%$ is found for stars with
$-2.5<[\Fe/\Hy]\le-2$. Considering the large uncertainties associated at
low metallicity with the initial distributions of masses and separations
in binary systems, the CEMP/VMP ratio predicted with our models is
consistent with the observed value, although the models predict many
more dwarf CEMP stars than are observed.

There are currently few measured orbital periods for CEMP stars, and
therefore it is difficult to provide strong constraints on our model of
angular-momentum loss and wind-accretion efficiency. Future observations
of the orbital periods of additional CEMP stars will allow us to gain insight into the
mass-transfer process at low metallicity. At present, the orbital
periods of most of our synthetic stars are on average ten times longer than the
typical observed periods. This indicates that our models should
produce more CEMP stars in binary systems below a few times $10^3$ days,
or, alternatively, that most CEMP stars have periods longer than $10^4$
days.

The population of synthetic CEMP stars qualitatively reproduces the
abundance distributions of carbon, sodium, and light-$s$ elements
observed in all CEMP stars. On the other hand, the observed
distributions of heavy-$s$ elements, lead, and europium are reproduced
only in $r$-normal CEMP-$s$ stars, whereas in CEMP-$s/r$ stars the
abundances of these elements are underestimated. The correlation between
the abundances of barium and europium observed in all CEMP stars
indicate that the enhancements in $s$- and $r$-elements are not
independent. This suggests that both the $s$- and $r$-processes may be
active in the same astrophysical site.

\begin{acknowledgements}

We thank the referee for her/his helpful and supportive comments on
our paper.
CA is grateful to P. van Oirschot for many useful discussions on the
stellar distribution in the Galactic halo and to the Netherlands Organisation 
for Scientific Research (NWO) for funding support under grant 614.000.901.
RJS is the recipient of a Sofja Kovalevskaja Award
from the Alexander von Humboldt Foundation. RGI is grateful to the Alexander von Humboldt Foundation
and the Science and Technology Facilities Council (STFC) for funding support.
AIK is supported through an Australian Research Council Future Fellowship (FT110100475).
TCB acknowledges partial support for this work from grants PHY 08-22648;
Physics Frontier Center/Joint Institute for Nuclear Astrophysics (JINA),
and PHY 14-30152; Physics Frontier Center/JINA Center for the Evolution
of the Elements (JINA-CEE), awarded by the US National Science
Foundation. Y.S.L. acknowledges partial support by the 2014 Research
Fund of the Chungnam National University.

\end{acknowledgements}

\bibliography{/home/carlo/work/articles/biblio/biblio}

\end{document}